\documentclass[aps,pra,twocolumn,showpacs,superscriptaddress,longbibliography]{revtex4-1}
\usepackage{graphicx} 
\usepackage{amsmath}
\usepackage{graphicx,epstopdf}
\usepackage{gensymb}
\newcommand{\be}{\begin{equation}}
\newcommand{\ee}{\end{equation}}
\newcommand{\bea}{\begin{eqnarray}}
\newcommand{\eea}{\end{eqnarray}}
\newcommand{\bse}{\begin{subequations}}
\newcommand{\ese}{\end{subequations}}

\newcommand{\av}{\mathbf a}
\newcommand{\bv}{\mathbf b}

\graphicspath{{../}}

\usepackage{color}
\usepackage[colorlinks,bookmarks=false,citecolor=darkblue,linkcolor=red,urlcolor=blue]{hyperref}

\definecolor{darkred}{rgb}{0.7,0.0,0.0}

\definecolor{darkblue}{rgb}{0,0.02,0.45}

\definecolor{darkgreen}{rgb}{0.02,0.45,0.0}

\definecolor{violet}{rgb}{0.8,0.2,0.6}

\begin{document}
\title{Quasi-one-dimensional magnetism in the spin-$\frac12$ antiferromagnet BaNa$_{2}$Cu(VO$_{4}$)$_{2}$}

\author{Sebin J. Sebastian}
\author{K. Somesh}
\affiliation{School of Physics, Indian Institute of Science
	Education and Research Thiruvananthapuram-695551, India}
\author{M. Nandi}
\affiliation{Ames Laboratory and Department of Physics and Astronomy, Iowa State University, Ames, Iowa 50011, USA}
\author{N. Ahmed}
\author{P. Bag}
\affiliation{School of Physics, Indian Institute of Science Education and Research Thiruvananthapuram-695551, India}
\author{M. Baenitz}
\affiliation{Max Planck Institute for Chemical Physics of Solids, Nöthnitzer Strasse 40, 01187 Dresden, Germany}
\author{B. Koo}
\affiliation{Max Planck Institute for Chemical Physics of Solids, Nöthnitzer Strasse 40, 01187 Dresden, Germany}
\author{J. Sichelschmidt}
\affiliation{Max Planck Institute for Chemical Physics of Solids, Nöthnitzer Strasse 40, 01187 Dresden, Germany}
\author{A. A. Tsirlin}
\email{altsirlin@gmail.com}
\affiliation{Experimental Physics VI, Center for Electronic Correlations and Magnetism, University of Augsburg, 86135 Augsburg, Germany}
\author{Y. Furukawa}
\affiliation{Ames Laboratory and Department of Physics and Astronomy, Iowa State University, Ames, Iowa 50011, USA}
\author{R. Nath}
\email{rnath@iisertvm.ac.in}
\affiliation{School of Physics, Indian Institute of Science Education and Research Thiruvananthapuram-695551, India}
\date{\today}
	
\begin{abstract}
We report synthesis and magnetic properties of quasi-one-dimensional spin-$\frac{1}{2}$ Heisenberg antiferromagnetic chain compound BaNa$_2$Cu(VO$_4$)$_2$. This orthovanadate has a centrosymmetric crystal structure, $C2/c$, where the magnetic Cu$^{2+}$ ions form spin chains. These chains are arranged in layers, with the chain direction changing by 62$\degree$ between the two successive layers. Alternatively, the spin lattice can be viewed as anisotropic triangular layers upon taking the inter-chain interactions into consideration. Despite this potential structural complexity, temperature-dependent magnetic susceptibility, heat capacity, ESR intensity, and NMR shift agree well with the uniform spin-$1/2$ Heisenberg chain model with an intrachain coupling of $J/k_{\rm B} \simeq 5.6$~K. The saturation field obtained from the magnetic isotherm measurement consistently reproduces the value of $J/k_{\rm B}$. Further, the $^{51}$V NMR spin-lattice relaxation rate mimics the 1D character in the intermediate temperature range, whereas magnetic long-range order sets in below $T_{\rm N} \simeq 0.25$~K. The effective interchain coupling is estimated to be $J_{\perp}/k_{\rm B} \simeq 0.1$~K. The theoretical estimation of exchange couplings using band-structure calculations reciprocate our experimental findings and unambiguously establish the 1D character of the compound. Finally, the spin lattice of BaNa$_2$Cu(VO$_4$)$_2$ is compared with the chemically similar but not isostructural compound BaAg$_2$Cu(VO$_4)_2$.
\end{abstract}
\maketitle
	
\section{Introduction}
The studies of low-dimensional and frustrated spin systems have contributed substantially in understanding the quantum phase transitions at low temperatures~\cite{Sachdev2007,Ramirez453}. In one-dimensional (1D) antiferromagnetic (AFM) spin-$1/2$ uniform Heisenberg chains, magnetic long-range-order (LRO) is forbidden at zero temperature as a result of enhanced quantum fluctuations, thereby exhibiting a gapless excitation spectrum and power-law decay of spin-spin correlations~\cite{Mermin1133}.
However, non-zero inter-chain interactions, inherent to real materials, lead to the formation of magnetic LRO at finite temperatures~\cite{Yasuda217201,Schulz2790}. On the other hand, the inter-chain interactions often create frustrated network between the chains that eventually prevents the system from achieving the conventional LRO but stabilizes different exotic states instead~\cite{Ramirez453,Greedan37,Kojima1787,Lancaster020410}. Further, competing interactions as realized in a set of compounds, add magnetic frustration in spin chains which along with quantum fluctuations host a multitude of intriguing magnetic ground states~\cite{Furukawa257205,Hase3651,Drechsler077202}. The transition-metal oxides offer nearly endless opportunities for realizing 1D spin chains with different types of exchange couplings, and may harbor wide varieties of exotic phases of matter.

\begin{figure*}
	\includegraphics[scale=0.2]{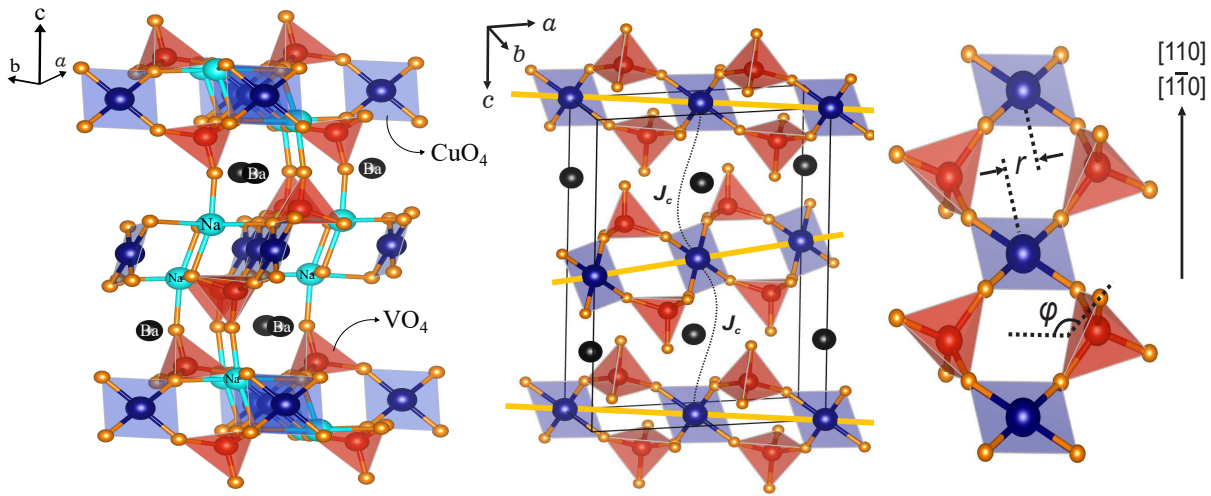}
	\caption{\label{Fig1} Left panel: crystal structure of BaNa$_2$Cu(VO$_4)_2$ showing corner-shared CuO$_4$ plaquettes and VO$_4$ tetrahedra forming layers of spin chains. The coupling of Na$^{1+}$ ions with the magnetic Cu$^{2+}$ ions is also shown. Middle panel: crystal structure of BaNa$_2$Cu(VO$_4)_2$ shown in a different orientation to visualize the spin chains running along the $[110]$ and $[1\bar 10]$ directions; black spheres show the Ba atoms, the Na atoms are omitted for clarity. Right panel: the structure of the single spin chain with the geometrical parameters $\varphi$ and $r$ that control the sign and strength of superexchange through the double bridges of the VO$_4$ tetrahedra.}
\end{figure*}	
Recently, synthesis and magnetic properties of a series of compounds $AA^{\prime} M$(VO$_4$)$_2$ ($A=$ Ba and Sr, $A^{\prime}=$ Na$_2$ and Ag$_2$, and $M=$ Mn, Ni, Co, Fe, and Cu) were reported. Despite some variations in their crystal structures, the magnetic model of anisotropic triangular lattice has been generally used to understand their magnetism~\cite{Amuneke2207,Moller214422,Nakayama116003,Reub6300,Sanjeewa2813,Amuneke5930,Lee224420}. BaAg$_2$Cu(VO$_4)_2$ stands as an exception in this series, because its crystal structure is triclinic (space group: $P\overline{1}$)~\cite{Amuneke2207}, and indeed microscopic analysis via density-functional band-structure calculations~\cite{Tsirlin014401} combined with resonance spectroscopy~\cite{Krupskaya759} revealed 1D magnetism with two dissimilar types of spin chains, one ferromagnetic and one antiferromagnetic, coexisting in the structure.

Here, we present for the first time the magnetic properties of BaNa$_2$Cu(VO$_4)_2$, another Cu$^{2+}$ member of the series~\cite{Von107}. Its structure features four equal Cu--Cu distances of 5.507\,\r A as well as two slightly longer distances of 5.686\,\r A, all in the $ab$ plane. This interaction geometry is a pre-requisite of the triangular-lattice scenario previously established for other members of the $AA^{\prime} M$(VO$_4$)$_2$ series. On the other hand, the square-planar oxygen coordination of Cu$^{2+}$ and the VO$_4$ bridges between such CuO$_4$ plaquette units may lead to one preferred direction for magnetic couplings in the $ab$ plane (Fig.~\ref{Fig1}). Interestingly, this preferred direction changes from $\av+\bv$ in one plane to $\av-\bv$ in the adjacent plane, thus leading to the formation of crossed spin chains arranged at $62^{\circ}$ relative to each other. This geometry resembles the crossed-chain magnetic model, where exotic ground states and potential spin-liquid behavior have been proposed theoretically~\cite{Starykh167203,Sindzingre174424,Brenig064420,Starykh094416,Bishop205122}.


Here, we use magnetization, heat capacity, electron spin resonance (ESR), and nuclear magnetic resonance (NMR) measurements, as well as complementary band-structure calculations to uncover magnetic interactions in BaNa$_2$Cu(VO$_4)_2$ and establish its microscopic magnetic model. Our data suggest the formation of uniform AFM spin chains with the exchange coupling $J/k_{\rm B} \simeq 5.6$~K, and the subsequent onset of magnetic LRO below $T_{\rm N} \simeq 0.25$~K. We suggest that this magnetic order can be driven by residual inter-chain couplings of $J_{\perp}/k_{\rm B} \simeq 0.1$~K that remain non-frustrated despite the crossed-chain structural geometry. Our results establish that the mere presence of spin chains arranged along two different directions is insufficient to reach the interesting physics of the crossed-chain model, and an additional condition for the lateral displacement of these chains has to be met experimentally.

\section{Methods}
\begin{figure}
	\centering
	\includegraphics[height=2.4in, width=3.4in]{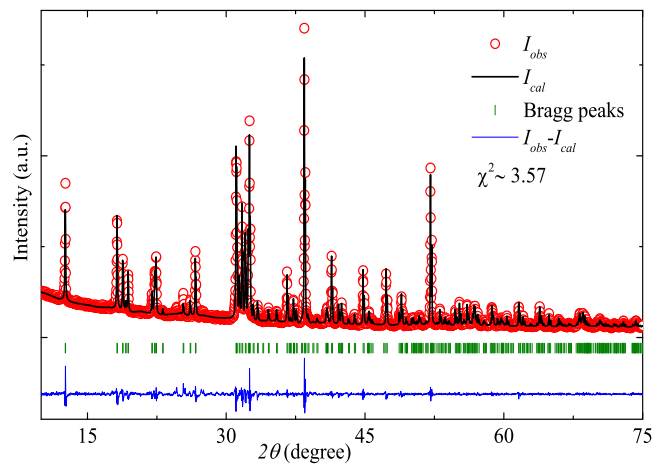}
	\caption{\label{Fig2} Powder XRD pattern of BaNa$_2$Cu(VO$_4$)$_2$ measured at $T = 300$~K. The circles are experimental data and the solid black line is the Le-Bail fit. The Bragg positions are indicated by green vertical lines and the bottom solid blue line indicates the difference between the experimental and calculated intensities.}
\end{figure}
Polycrystalline sample of BaNa$_2$Cu(VO$_4$)$_2$ was prepared by the usual solid-state reaction method. Initially, the reactants Na$_{2}$CO$_{3}$ (Aldrich, 99.995\%), BaCO$_{3}$ (Aldrich, 99.995\%), CuO (Aldrich, 99.999\%), and V$_{2}$O$_{5}$ (Aldrich, 99.995\%) were mixed in proper molar ratios, thoroughly ground, and then pressed into pellets. The pellets were sintered in an alumina crucible at 500~$^{0}$C for three days in air with several intermediate grindings. The phase purity of the sample was confirmed from the powder x-ray diffraction (XRD) performed at room temperature. For the powder XRD experiment, a PANalytical powder diffractometer with Cu\textit{K}$_{\alpha}$ radiation ($\lambda_{\rm avg} \simeq 1.54182$~{\AA}) was used. Le-Bail analysis of the powder XRD pattern was performed using the \verb"FullProf" software package~\cite{Rodriguez55}. Figure~\ref{Fig2} displays the room-temperature powder XRD data along with the fit. The structural parameters given in Ref.~\cite{Von107} were used as the initial parameters. The goodness-of-fit was found to be $ \chi^{2} \simeq 3.57$. The obtained lattice parameters are $a = 9.4379(1)$~{\AA}, $b = 5.6926(1)$~{\AA}, $c = 14.0519(1)$~{\AA}, and $\beta = 92.3434(8)^{\circ}$ and the unit cell volume $V_{cell}\simeq 754.34$~{\AA}$^{3}$, which are in close agreement with the previous report~\cite{Von107}.

Magnetization ($M$) measurements were performed as a function of temperature (0.48~K~$\leq T \leq 380$~K) and magnetic field ($0 \leq H \leq 14$~T) using a Superconducting Quantum Interference Device (SQUID, Quantum Design) magnetometer and a Physical Property Measurement System (PPMS, Quantum Design). The SQUID enabled us to measure magnetization down to 0.48~K with a $^{3}$He attachment. High-field magnetization up to 14~T were measured using PPMS. Heat capacity ($C_{\rm p}$) was measured as a function of $T$ (0.4~K~$\leq T \leq 200$~K) on a sintered pellet using the thermal relaxation method in PPMS. The temperature down to 0.4~K was achieved using a $^{3}$He attachment to the PPMS.

The ESR experiments were performed on the powder sample with a standard continuous-wave spectrometer in the temperature range 2.5~K$\leq T \leq 300$~K. As a function of external magnetic field $B$, the resonance shows up as an absorbed power $P$ of a transversal magnetic microwave field ($\nu \simeq 9.4$~GHz, X-band). In order to improve the signal-to-noise ratio, a lock-in technique was used by modulating the applied field, which yields the derivative of power absorption $dP/dB$ as a function of $B$. By using the resonance condition $g=\frac{h\nu}{\mu_{\rm B}H_{\rm res}}$, where $h$ is the Planck's constant, $\mu_{\rm B}$ is the Bohr magneton, $\nu$ is the resonance frequency, and $H_{\rm res}$ is the corresponding resonance field, the $g$-value was obtained.

The pulsed NMR measurements were performed on both $^{23}$Na (nuclear spin $I=3/2$ and gyromagnetic ratio $\gamma = 11.26$~MHz/T) and $^{51}$V ($I=7/2$ and $\gamma = 11.19$~MHz/T) nuclei in the temperature range 0.044~K$\leq T \leq 200$~K. For measurements above 2~K a $^{4}$He cryostat (Oxford Instrument) with a field-sweep superconducting magnet was used, while for measurements in the low-temperature range (0.044~K$\leq T \leq 2$~K), a $^{3}$He/$^{4}$He dilution refrigerator (Kelvinox, Oxford Instruments) with a field sweep magnet was used. All the NMR measurements were carried out in a radio frequency of 77~MHz. The NMR spectra were measured as a function of temperature $T$ by sweeping the magnetic field at a constant radio frequency of 77~MHz. The NMR shift was calculated for both $^{23}$Na and $^{51}$V nuclei as $K(T)$ = [$H_{\rm ref}-H(T)$]/$H(T)$, where $H$ is the resonance field for $^{23}$Na and $^{51}$V and $H_{\rm ref}$ is the resonance field of the non-magnetic reference sample. The spin-lattice relaxation rate $1/T_{1}$ was measured by the conventional single saturation pulse method.

Density-functional (DFT) band-structure calculations were performed in the \texttt{FPLO} code~\cite{fplo} using the structural parameters from Ref.~\cite{Von107} and local-density approximation (LDA) for the exchange-correlation potential~\cite{Perdew13244}. Exchange parameters of the spin Hamiltonian
\begin{equation}
	\mathcal H=\sum_{\langle ij\rangle} J_{ij}\mathbf S_i\mathbf S_j
	\label{Eq1}
\end{equation}
with $S=\frac12$ and the summation over atomic pairs $\langle ij\rangle$, were extracted via two complementary procedures. First, band structure obtained on the LDA level was mapped onto a tight-binding model for the half-filled $d_{x^2-y^2}$ orbitals of Cu$^{2+}$ as the magnetic ion. Squared hopping parameters $t_i$ of this tight-binding model are proportional to AFM contributions to the exchange, $J_i^{\rm AFM}=4t_i^2/U_{\rm eff}$, where $U_{\rm eff}$ is the effective on-site Coulomb repulsion. Alternatively, full exchange couplings $J_i$ comprising both FM and AFM contributions are extracted by a mapping procedure~\cite{Xiang224429} from total energies of magnetically ordered states calculated on the DFT+$U$ level, with correlation effects in the Cu $3d$ shell modeled by the on-site Coulomb repulsion $U_d=6$\,eV, Hund's exchange $J_d=1$\,eV, and around-mean-field flavor of the double-counting correction~\cite{Janson144423,Tsirlin014401}. The $k$ mesh with up to 150 points in the symmetry-irreducible part of the first Brillouin zone was used. 

Field-dependent magnetization and magnetic specific heat of a uniform spin-$\frac12$ chain were obtained from quantum Monte-Carlo simulations for $L=32$ finite lattices with periodic boundary conditions. The \texttt{loop}~\cite{loop} and \texttt{dirloop\_sse}~\cite{dirloop} algorithms of the \texttt{ALPS} simulation package~\cite{alps} were used.

\section{Results and Discussion}
\subsection{Magnetization}
\begin{figure}
\includegraphics[height=2.4in, width=3.4in]{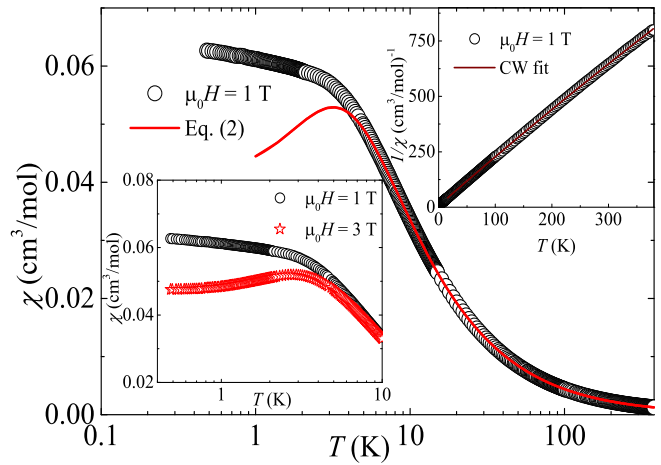}
\caption{\label{Fig3} $\chi$ of polycrystalline BaNa$_2$Cu(VO$_4$)$_2$ sample as a function of temperature in an applied field $\mu_0H = 1$~T. The
solid line is the fit using Bonner-Fisher model [Eq.~\eqref{Eq2}] for
uniform Heisenberg spin-$1/2$ chain. Upper inset: inverse
susceptibility $1/\chi$ vs $T$ and the solid line represents the CW fit,
as discussed in the text. Lower inset: the low-temperature $\chi(T)$ measured in two different fields $\mu_0H = 1$~T and 3~T.}
\end{figure}
Temperature-dependent magnetic susceptibility $\chi(T) (= M/H$) of the polycrystalline Na$_{2}$BaCu(VO$ _{4} $)$ _{2} $ sample measured in two different applied fields $H = 1$~T and 3~T is depicted in the Fig.~\ref{Fig3}. The most significant feature in the $\chi(T)$ curve is the presence of a broad maximum at 3~K, signaling a crossover to an AFM short-range ordered state, typical for low-dimensional spin systems~\cite{BonnerA640,Eggert332}. This broad maximum is more pronounced in the 3~T data shown in the lower inset of Fig.~\ref{Fig3}. No anomaly indicative of the potential LRO could be seen down to 0.48~K. 	

The preliminary analysis was done by fitting the $\chi(T)$ data using the Curie-Weiss (CW) law, $\chi(T) = \chi_{0} + C/(T+\theta_{\rm CW}$), where $\chi_{0}$ is the temperature-independent susceptibility, $C$ is the Curie constant, and $\theta_{\rm CW}$ is the characteristic CW temperature. The fit shown in the upper inset of Fig.~\ref{Fig3} in the high-temperature regime ($T \geq 16$~K) yields the following parameters: $\chi_{0} \simeq 7.9288 \times 10^{-5}$~cm$^{3}$/mol, $C \simeq 0.445$~cm$^{3}$K/mol, and $\theta_{\rm CW} \simeq 3$~K. In order to estimate the Van-Vleck paramagnetic susceptibility ($\chi_{\rm VV}$), which arises from the second-order contribution to free energy in the presence of magnetic field, core diamagnetic susceptibility $\chi_{\rm core}$ of Na$_{2}$BaCu(VO$ _{4} $)$ _{2} $ was calculated to be $-1.57 \times 10^{-4}$~cm$^{3}$/mol by summing the core diamagnetic susceptibilities of individual ions Na$^{+}$, Ba$^{2+}$, Cu$^{2+}$, V$^{5+}$, and O$^{2-}$~\cite{Selwood2013,Mendelsohn1130}. Subsequently, $\chi_{\rm VV}$ was obtained by subtracting $\chi_{\rm core}$ from $\chi_0$ to be $\sim 2.36 \times 10^{-4}$~cm$^{3}$/mol, which is close to the values reported for other cuprates~\cite{Motoyama3212,Nath174436,Ahmed214413} and consistent with tetragonal crystal-field splitting at the Cu$^{2+}$ site with the square-planar oxygen coordination~\cite{Takigawa1989}.

From the Curie constant $C$, the effective moment is calculated using the relation $\mu_{\rm eff} = \sqrt{3k_{\rm B}C/N_{\rm A}}$ to be $\simeq 1.88$~$\mu_{\rm B}$, where $k_{\rm B}$ is the Boltzmann constant,  $\mu_{\rm B}$ is the Bohr magneton, and $N_{\rm A}$ is the Avogadro's number. For a spin-$\frac{1}{2}$ system, the spin-only effective moment is expected to be $\mu_{\rm eff} = g\sqrt{S(S+1)}\mu_{\rm B} \simeq 1.73$~$\mu_{\rm B}$, assuming Land\'e $g$-factor $g = 2$. However, our experimental value of $\mu_{\rm eff} \simeq 1.88$~$\mu_{\rm B}$ corresponds to a $g$-factor of $g \simeq 2.17$, which is consistent with the ESR experiments discussed later. The positive value of $\theta_{\rm CW}$ suggests that the dominant exchange interactions between the Cu$^{2+}$ ions are AFM in nature.
	
In order to estimate the exchange coupling between the Cu$^{2+}$ ions, we decomposed $\chi(T)$ into three components,
\begin{equation}
\chi(T)=\chi_0+\frac{C_{\rm imp}}{T} + \chi_{\rm spin}(T).
\label{Eq2}
\end{equation}
Here, the second term is the Curie law, which accounts for the paramagnetic contributions from impurity spins and/or defects, and $\chi_{\rm spin}(T)$ is the intrinsic spin susceptibility. This last term can be chosen in different forms depending on the underlying magnetic model. The best fit was achieved with the spin-chain model, which is further supported by the specific-heat data (Sec.~\ref{sec:heat}) and \textit{ab initio} calculations (Sec.~\ref{sec:model}).

The susceptibility of a spin-$\frac{1}{2}$ uniform Heisenberg AFM chain takes the form
\begin{equation}
	\chi_{\rm spin} =\frac{N_{A}\mu_{B}^{2}g^{2}}{k_{B}T} \frac{0.25+0.0775x+0.0752x^{2}}{1+0.993x+0.1721x^{2}+0.7578x^{3}},
	\label{Eq3}
\end{equation}
with $x=\lvert J \rvert/k_{\rm B}T$~\cite{BonnerA640}. This is simply a high-temperature series expansion (HTSE) valid in the regime $k_{\rm B}T/J \geq 0.5$. The solid line in Fig.~\ref{Fig3} represents the best fit of the $\chi(T)$ data above 4~K by Eq.~\eqref{Eq2}. The following parameters were obtained: $\chi_{0} \simeq 1.44 \times 10^{-4}$~cm$^{3}$/mol, $C_{\rm imp}\simeq 0.0258$~cm$^{3}$K/mol, $g \simeq 2.13$, and the dominant intra-chain AFM exchange coupling $J/k_{\rm B} \simeq 5.6$~K. From the value of $C_{\rm imp}$, the sample was found to contain $\sim 6$\% spin-$\frac{1}{2}$ impurities/defects. At temperatures below 1\,K, this impurity contribution becomes dominant and causes the reduction in the susceptibility with the applied field, even though $\chi_{\rm spin}(T)$ should increase when the field is applied~\cite{klumper1998}.

\begin{figure}
	\includegraphics[height=2.4in, width=3.4in] {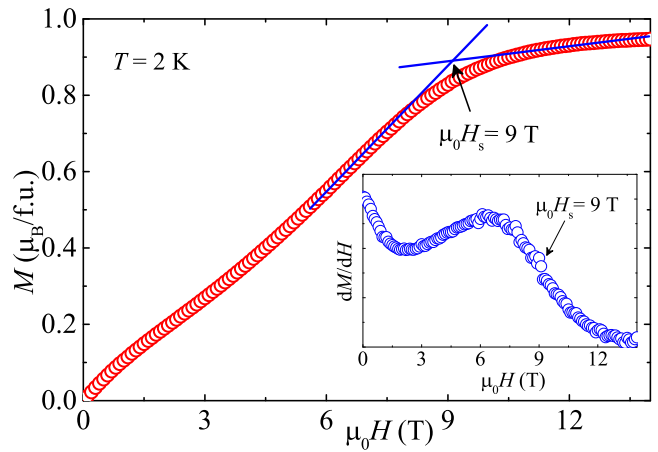}
	\caption{\label{Fig4} Magnetization ($M$) vs $H$ measured at $T= 2$~K. Inset: $dM/dH$ vs $H$ to highlight the saturation field $H_{\rm s}$.}
\end{figure}
The magnetic isotherm at $T = 2$~K upto 14~T is shown in Fig.~\ref{Fig4}. $M$ increases almost linearly with $H$ but with a small curvature. It develops a tendency of saturation above 9~T. A more accurate value of the saturation field $H_{\rm s} \simeq 9$~T was found by drawing tangential at the curvature (see Fig.~\ref{Fig4}). The field derivative of the $M$ vs $H$ plot also implies $H_{\rm s} \simeq 9$~T (see the inset of Fig.~\ref{Fig4}). For a spin-$1/2$ Heisenberg AFM chain, the saturation field is directly proportional to the intra-chain exchange coupling as $H_{\rm s}=2J_{\rm 1D}(k_B/g\mu_B)$~\cite{Lebernegg174436}. Using the value of $J/k_{\rm B} \simeq 5.6$~K, the saturation field is calculated to be $H_{\rm s} \simeq 8.34$~T, which matches well with the experimental value, confirming the dominant 1D character of the compound.

\subsection{ESR}
\begin{figure}
	\includegraphics[] {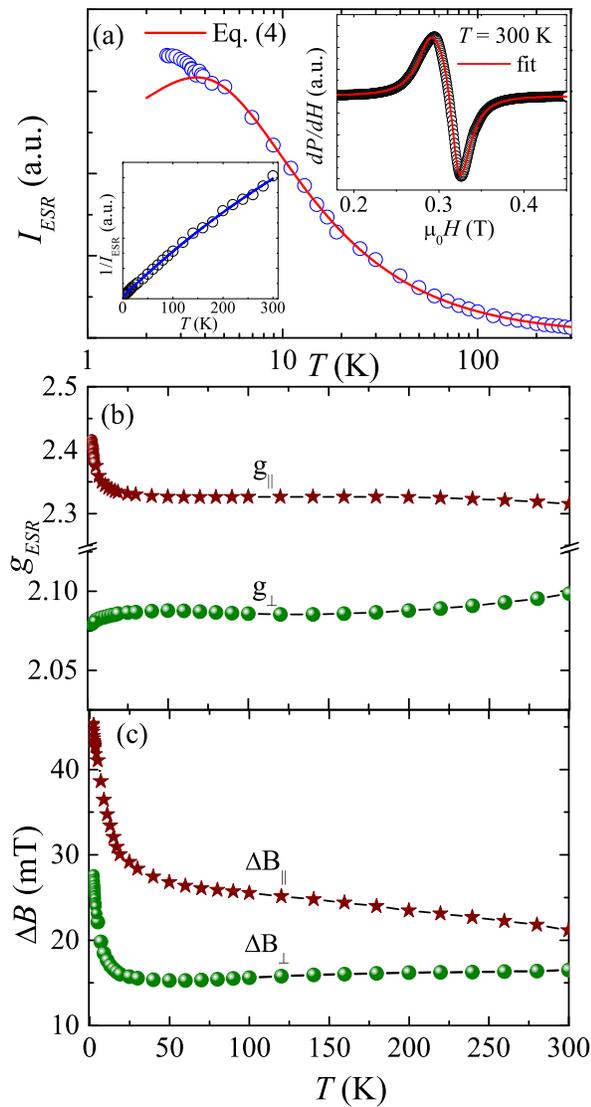}
	\caption{\label{Fig5}(a) Integrated ESR intensity vs temperature and the solid line represents the fit as described in the text. Inset: ESR spectrum at room temperature measured at a microwave frequency of 9.4~GHz together with the powder-averaged Lorentzian fit (solid line). (b) Temperature variation of the $g$ values (both perpendicular and parallel components) obtained from the Lorentzian fit. (c) Temperature-dependent ESR linewidth $\Delta B $ (both perpendicular and parallel components).}
\end{figure}
ESR experiment was performed on the powder sample and the results are shown in Fig.~\ref{Fig5}. The inset of Fig.~\ref{Fig5}(a) depicts a typical ESR powder spectrum at 300~K. The uniaxial $g$~factor anisotropy was obtained by fitting the spectra using the powder-averaged Lorentzian line. The fit of the spectral at room temperature yields the anisotropic $g$-tensor components $g_{\parallel}$$\simeq 2.315$ and $g_{\perp}$$\simeq 2.098$. From these values, the average $g$-value was calculated as $g=[(g_{\parallel}+2g_{\perp})/3] \simeq 2.17$~\cite{Abragam2012electron}. This value is slightly larger ($\Delta g/g \simeq 0.085$) compared to the free electron value ($g=2$), typical for Cu$^{2+}$ based oxides~\cite{Kochelaev4274,Nath014407}.
The integrated ESR intensity ($I_{\rm ESR}$) obtained from the above fit is plotted as a function of temperature in Fig.~\ref{Fig5}(a). It shows similitude with the $\chi(T)$ behavior, which traces a broad maximum at around $T^{\max}_{\rm ESR}$ $\simeq 3.7$~K. Indeed, the $I_{\rm ESR}$ vs $\chi$ plot with temperature as an implicit parameter follows a straight line down to $\sim 5$~K (not shown). The variation of $g$ with respect to temperature is shown in Fig.~\ref{Fig5}(b). Both the components of $g$ were found to be almost temperature-independent at high temperatures ($T \geq 20$~K). However, below 20~K a weak deviation from the room-temperature values is observed.

In order to estimate the exchange coupling, $I_{\rm ESR}(T)$ was fitted by
\begin{equation}
I_{\rm ESR}(T) = A + B \chi_{\rm spin}(T).
\label{Eq4}
\end{equation}
Here, $A$ and $B$ are arbitrary constants, and $\chi_{\rm spin}$ is given by Eq.~\eqref{Eq3}. Our fit (see Fig.~\ref{Fig5}) in the high-temperature regime ($T \geq 5$~K) produced $J/k_{\rm B} \simeq 5.55~K$. This value of $J/k_{\rm B}$ is close to the one obtained from the $\chi(T)$ analysis. During the fit, the value of $g$ was kept constant to 2.17, as obtained above. We have also fitted the $1/I_{\rm ESR}$ data in the high-temperature regime ($T \geq 10$~K) using the relation $I_{\rm ESR} = M+N/(T+\theta_{\rm CW})$ where $M$ and $N$ are arbitrary constants. As shown in the lower inset of Fig.~\ref{Fig5}(a), the fit returns $\theta_{\rm CW}\simeq 3.9~K $, which is in good agreement with the value obtained from the $\chi^{-1}(T)$ analysis. 

The temperature-dependent ESR linewidth, or equivalently the half-width at half maximum of the ESR absorption signal, is presented in Fig.~\ref{Fig5}(c). Both the parallel ($\Delta B_{\parallel}$) and perpendicular ($\Delta B_{\perp}$) components of the ESR line width follow the general trend, commonly observed in most of the low-dimensional spin systems~\cite{Ivanshin064404,Sichelschmidt75}. The rapid increase/divergence below $\sim 25$~K indicates the growth of strong spin correlations at low temperatures as the system approaches the magnetic LRO state.

\subsection{Heat Capacity}
\label{sec:heat}
\begin{figure}[htb!]
	\includegraphics[height=4.6in, width=3.5in] {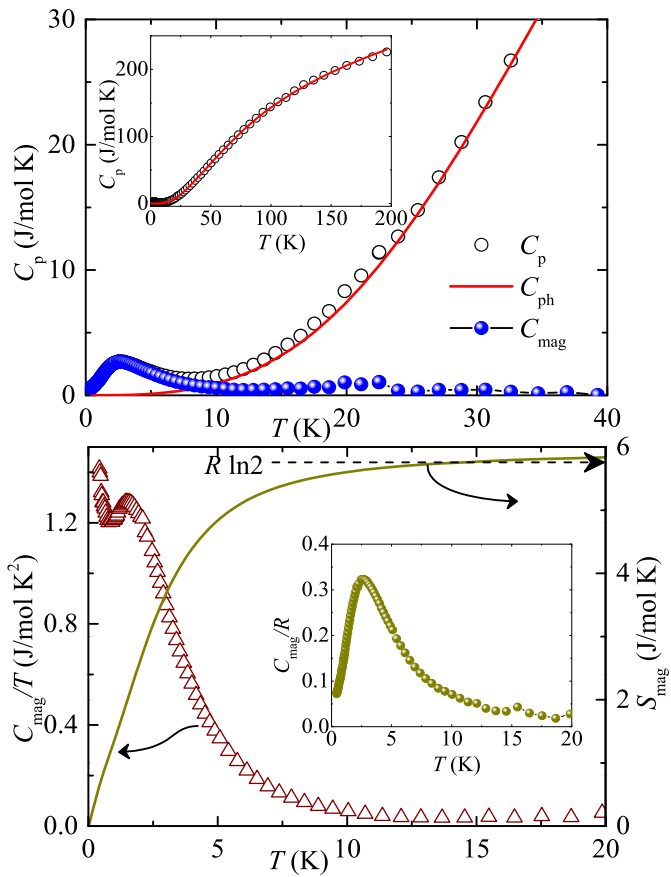}
	\caption{\label{Fig6} Upper panel: Heat capacity ($C_{\rm p}$) vs $T$ in zero applied field. The solid line denotes the phonon contribution to the heat capacity $C_{\rm ph}$ using the Debye-Einstein fit. The blue solid spheres indicate the magnetic contribution to the heat capacity $C_{\rm mag}$. Inset: $C_{\rm p}$ vs $T$ in the whole measured temperature range along with the Debye-Einstein fit. Lower panel: The left $y$-axis shows $C_{\rm mag}/T$ and the right $y$-axis shows the magnetic entropy $S_{\rm mag}$ vs $T$. Inset: $C_{\rm mag}/R$ vs $T$.}
\end{figure}
Temperature-dependent heat capacity $C_{\rm p}$ of the polycrystalline sample is shown in the upper panel of Fig.~\ref{Fig6}. In magnetic insulators, the two major contributions to $C_{\rm p}$ are from phonon and magnetic parts. At high temperatures, $C_{\rm p}(T)$ is dominated by the phonon part, while at low temperatures it is dominated by the magnetic part. Our experimental $C_{\rm p}$ data exhibit a pronounced broad maximum at $T \simeq 1.52$~K indicative of the low-dimensional short-range order and also reflects the dominant magnetic contribution at low temperatures. In order to estimate the magnetic contribution to the heat capacity $C_{\rm mag}$, we proceed as follows. First we approximate the lattice contribution $C_{\rm ph}$ by fitting the high-temperature data by a linear combination of one Debye and two Einstein terms (Debye-Einstein model) as~\cite{Kittelc2005,Caslin014412}
\begin{equation}
	C_{\rm ph}(T)=f_{\rm D}\,C_{\rm D}(\theta_{\rm D},T)+\sum_{i = 1}^{2}g_{i}\,C_{{\rm E}_i}(\theta_{{\rm E}_i},T).
	\label{Eq5}
\end{equation}
The first term in Eq.~\eqref{Eq5} is the Debye term,
\begin{equation}
	C_{\rm D} (\theta_{\rm D}, T)=9nR\left(\frac{T}{\theta_{\rm D}}\right)^{3} \int_0^{\frac{\theta_{\rm D}}{T}}\frac{x^4e^x}{(e^x-1)^2} dx.
	\label{Eq6}
\end{equation}
Here, $x=\frac{\hbar\omega}{k_{\rm B}T}$, $\omega$ is the vibration frequency, $R$ is the universal gas constant, and $\theta_{\rm D}$ is the characteristic Debye temperature.
The second term in Eq.~\eqref{Eq5} is a combination of the Einstein terms that are usually responsible for flat optical modes in the phonon spectrum,
\begin{equation}
C_{\rm E}(\theta_{\rm E}, T) = 3nR\left(\frac{\theta_{\rm E}}{T}\right)^2 
\frac{e^{\,\theta_{\rm E}/T}}{\left(e^{\,\theta_{\rm E}/T}-1\right)^2}.
\label{Eq7} 
\end{equation}
Here, $\theta_{\rm E}$ is the characteristic Einstein temperature. The coefficients $f_{\rm D}$, $g_1$, and $g_2$ are the weight factors, which take into account the number of atoms per formula unit ($n$) and are conditioned such that at high temperatures the Dulong-Petit value of $3nR$ is satisfied. The $C_{\rm p}(T)$ data above $\sim 15$~K were fitted by Eq.~\eqref{Eq5} and the obtained parameters are $f_{\rm D} \simeq 0.34$, $g_1 \simeq 0.35$, and $g_2 \simeq 0.31$, $\theta_{\rm D} \simeq 214$~K, $\theta_{{\rm E}_1} \simeq 356$~K, and $\theta_{{\rm E}_2} \simeq 897$~K. Further Einstein terms beyond $\theta_{{\rm E}_2}$ rendered the fit unstable. The fit itself is phenomenological in nature, although one may tentatively associate $\theta_{\rm D}$ with low-energy vibrations of heavier atoms (Ba, Cu, and V) that constitute 28.5\,\%, about $\frac13$ of the atomic species in BaNa$_2$Cu(VO$_4$)$_2$. The lower Einstein temperature $\theta_{{\rm E}_1}$ may correspond to Na atoms and two apical oxygens of the VO$_4$ tetrahedra (altogether 6 atoms per formula unit), whereas $\theta_{{\rm E}_2}$ reflects higher-energy vibrations of the remaining four oxygens that are bound to V and Cu at the same time.

The high-$T$ fit was extrapolated down to low temperatures and $C_{\rm mag}(T)$ was estimated by subtracting $C_{\rm {ph}}(T)$ from $C_{\rm p}(T)$ [see Fig.~\ref{Fig6} (upper panel)]. $C_{\rm mag}(T)/T$ was plotted as a function of temperature in the lower panel of Fig.~\ref{Fig6}. The broad maximum corresponding to the short-range order is apparent at $T \simeq 1.52$~K. At low temperatures, $C_{\rm mag}(T)/T$ shows a rapid increase, which could be related to the onset of magnetic LRO below 0.4~K. The magnetic entropy was calculated as $S_{\rm{mag}}(T) = \int_{\rm 2\,K}^{T}\frac{C_{\rm {mag}}(T')}{T'}dT'$, which yields $S_{\rm mag} \simeq 5.83$~J/mol~K at 20~K (see the lower panel of Fig.~\ref{Fig6}). This value is close to the expected magnetic entropy for spin-$\frac{1}{2}$: $S_{\rm mag}=R\ln 2=5.76$~J/mol~K. 

In the inset of the lower panel of Fig.~\ref{Fig6}, $C_{\rm mag}/R$ is plotted against $T$. The peak of $C_{\rm mag}/R$ can be used to discriminate between different microscopic scenarios. Its height depends on the nature of the underlying spin lattice~\cite{Bernu134409}. Our experimental peak value of $C_{\rm mag}/R \simeq 0.323$ fits well to the aforementioned 1D scenario, suggesting that VO$_4$ bridges choose the direction of spin chains. Alternatively, four shortest Cu--Cu contacts of 5.507\,\r A could cause interactions of equal strength and form a 2D square-lattice interaction topology that should manifest itself by a much higher peak with $C_{\rm mag}/R\simeq 0.47$. On the other hand, the triangular-lattice scenario would reduce the peak value to $C_{\rm mag}/R\simeq 0.22$, lower than seen experimentally. We thus conclude that our specific-heat data favor the spin-chain scenario for BaNa$_2$Cu(VO$_4)_2$.



\subsection{$^{23}$Na and $^{51}$V NMR}
NMR is a potent tool to study the static and dynamic properties of spin systems. In Na$_{2}$BaCu(VO$_{4}$)$_{2}$, the $^{23}$Na and $^{51}$V nuclei are hyperfine-coupled to the magnetic Cu$^{2+}$ ions along the spin chains. Therefore, the low-lying excitations of Cu$^{2+}$ spins can be probed by means of $^{23}$Na and $^{51}$V NMR measurements.

\begin{figure}
	\includegraphics[height=4.0in, width=3.5in] {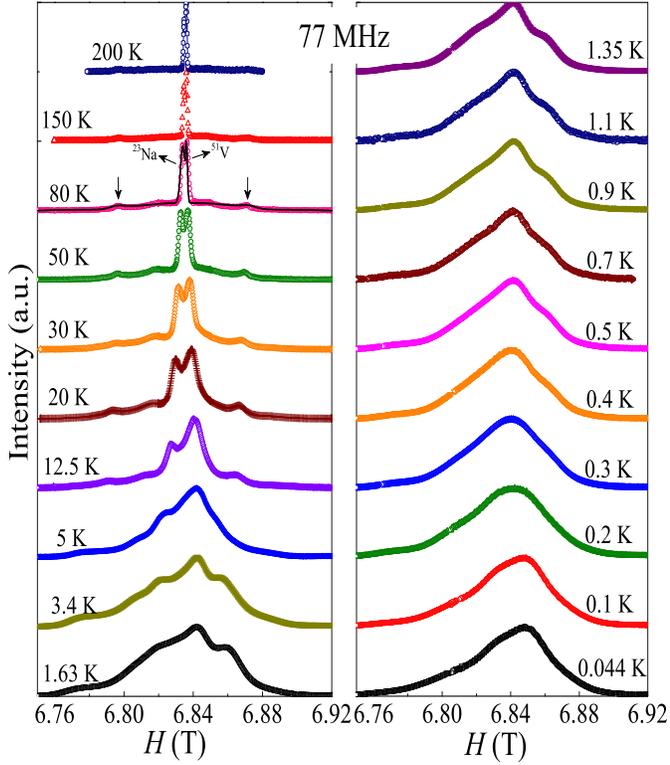}
	\caption{\label{Fig7} Field-sweep NMR spectra of the polycrystalline Na$_{2}$BaCu(VO$_{4}$)$_{2}$ sample, measured at 77~MHz as a function of temperature. The spectral lines corresponding to $^{23}$Na and $^{51}$V nuclei for $T = 80$~K are marked by arrows. The solid line is the simulated spectrum.}
\end{figure}
The quadrupole nuclei $^{23}$Na ($I = 3/2$) and $^{51}$V ($I = 7/2$) are in a non-cubic symmetry that may produce an asymmetric charge distribution and hence electric field gradient (EFG). Therefore, the four-fold and eight-fold degeneracies of the $I = 3/2$ and $I = 7/2$ spins, respectively, are lifted partially due to the interaction between the nuclear quadrupole moment ($Q$) and the surrounding EFG.
In this case, the nuclear spin Hamiltonian is a sum of the Zeeman and quadrupolar interaction terms~\cite{Curro026502,Slichter1992},
\begin{equation}
\mathcal{H} = -\gamma \hbar {\hat I} H(1+K) + \frac{h\nu_Q}{6}[(3\hat{I}_{z}^2-\hat{I}^2)+\eta(\hat{I}_{x}^2-\hat{I}_{y}^2)].
\label{Eq8}
\end{equation}
Here, the nuclear quadrupole resonance (NQR) frequency is defined as $\nu_{\rm Q}= \frac{3e^2qQ}{2I(2I-1)h}$, $e$ is the electron charge, $\hbar~(=h/2\pi)$ is the Planck's constant, $H$ is the applied field along $\hat{z}$, $K$ is the magnetic shift due to hyperfine field at the nuclear site, $V_{\alpha \beta}$ are the components of the EFG tensor, $eq = V_{z z}$ is the largest eigenvalue or principal component of the EFG, and $\eta=|V_{xx}-V_{yy}|/V_{zz}$ is the EFG asymmetry (here, the principal axes of EFG are chosen such that $|V_{zz}| \geq |V_{yy}| \geq |V_{xx}|$.). Experimentally, the transitions can be observed at the frequency $\nu_z = \nu_Q \sqrt{1+\eta^2/3}$.

The principal axes $\{x,y,z \}$ of the EFG tensor are defined by the local symmetry of the crystal structure. Consequently, the corresponding resonance frequency to any nuclear transition will have strong dependence on the direction of the applied field with respect to the crystallographic axes. For a site with axial symmetry ($\eta = 0$), there will be $2I-1$ quadrupolar resonances at frequencies $n\nu_{\rm Q}$, where $n=$ 1, ....$2I-1$. When $\eta > 0$, the resonances are not equally spaced. The EFG is fully characterized by the parameters $ \nu_{\rm z} $, $\eta$, and $\hat{z}$, where $\hat{z}$ is the unit vector in the direction of the principal axis of the EFG with the largest eigenvalue. When the Zeeman term dominates over the quadrupole term, first-order perturbation theory is enough for describing the system. In such a scenario, for a quadrupole nucleus, equally spaced satellite peaks should appear on either side of the central peak separated by $\nu_Q$~\cite{Lang094404}.

The NMR spectra as a function of temperature measured by sweeping the magnetic field at 77~MHz are presented in Fig.~\ref{Fig7}. Since $^{23}$Na and $^{51}$V nuclei have nearly the same $\gamma$ values, one expects their spectral lines to appear very close to each other. Further, $^{23}$Na and $^{51}$V are quadrupolar nuclei with nuclear spins $I = 3/2$ and $7/2$, respectively and the transitions with $\Delta m = \pm 1$ are expected between the energy levels. Therefore, one would anticipate three NMR lines for $^{23}$Na: one central line corresponding to $I_z =+1/2\longleftrightarrow-1/2$ and two equally spaced satellite lines corresponding to $I_z =\pm3/2\longleftrightarrow\pm1/2 $ and seven NMR lines for $^{51}$V: the central line being $I_z =+1/2\longleftrightarrow-1/2$ and the satellite lines $I_z = \pm1/2 \longleftrightarrow\pm 3/2 \longleftrightarrow\pm 5/2 \longleftrightarrow\pm 7/2 $. Indeed, at high temperatures, we observed two sharp and prominent peaks at the resonance field position and two satellite peaks on either side of those. The central peak towards the low-field side is identified to be the signal coming from the $^{23}$Na nuclei, while the one towards the high-field side appears to be the $^{51}$V peak. In addition to the central peaks, two satellite peaks correspond to the $^{23}$Na line. At high temperatures, the NMR spectra are found to be narrow and one can distinguish the $^{23}$Na and $^{51}$V signals. As the temperature is lowered, the line broadens asymmetrically and the central lines shift weakly with temperature. No abrupt line broadening was noticed down to 44~mK, which may signal the absence of magnetic LRO~\cite{Ranjith014415}.

\begin{figure}
	\includegraphics[height=4.6in, width=3.5in] {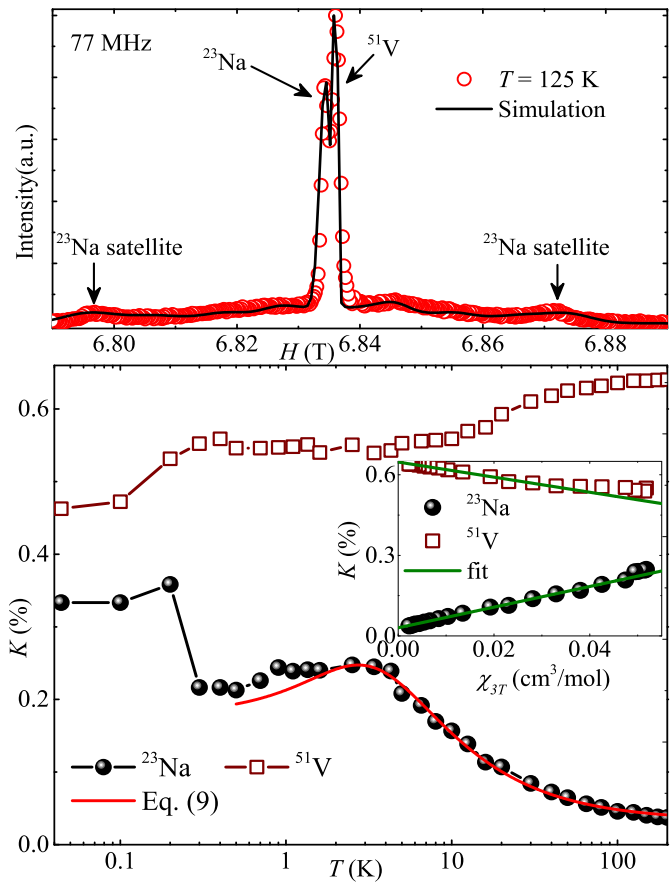}
	\caption{\label{Fig8} Upper panel: NMR spectra at $T = 125$~K showing the $^{23}$Na and $^{51}$V central lines, with the downward arrows pointing to the $^{23}$Na satellites. The solid line is the simulation of the spectra assuming the superposition of the $^{23}$Na and $^{51}$V signals. Lower panel: temperature-dependent NMR shift $K$ as a function of temperature for $^{23}$Na and $^{51}$V, measured at 77~MHz. Solid line is the fit using Eq.~\eqref{Eq9}. Inset: NMR shift vs $\chi$ measured at 3~T. Solid lines are the linear fits.}
\end{figure}
The spectra were fitted assuming the superposition of $^{23}$Na and $^{51}$V signals. The spectral fit at $T = 125$~K is presented in the upper panel of Fig.~\ref{Fig8}, where $^{23}$Na and $^{51}$V lines and their satellites are marked by arrows. The obtained fitting parameters are $K \simeq 0.0345$\% (isotropic shift), $\eta = 0$ (asymmetry parameter), and $\nu_Q \simeq 0.92$~MHz (NQR frequency) for $^{23}$Na and $K \simeq 0.627$\%, $\eta = 0$, and $\nu_Q \simeq 0.234$~MHz for $^{51}$V. The quadrupole frequency is found to be almost constant with temperature down to 1.5~K, which essentially excludes the possibility of any structural distortion in the studied compound.

The NMR shift $K(T)$ for both $^{23}$Na and $^{51}$V lines obtained from the spectral fits is plotted in the lower panel of Fig.~\ref{Fig8}. The temperature-dependent $^{23}$Na shift [$^{\rm 23}K(T)$] is found to have a broad maximum at around 3~K, similar to the $\chi(T)$ data.
As $K(T)$ in an intrinsic measure of the spin susceptibility $\chi_{\rm spin}$, one can write the linear relation
\begin{equation}
	K(T) = K_{0}+\frac{A_{\rm hf}}{N_{A}\mu_{B}}\chi_{\rm spin},
	\label{Eq9}
\end{equation}
where $K_{0}$ is the temperature-independent chemical shift and the proportionality constant $A_{\rm hf}$ is the hyperfine coupling between the probed nuclei and the electron spins.

From Eq.~\eqref{Eq9}, $A_{\rm hf}$ can be calculated by taking the slope of the linear $K$ vs $\chi$ plot (inset of Fig.~\ref{Fig8}) with temperature as an implicit parameter. In the case of $^{23}$Na, the data for $T\geq 5$~K were fitted well by a linear function, and the slope of the fit yields $^{23}A_{\rm hf} \simeq 0.021$~T/$\mu_{\rm B}$. Similarly, for $^{51}$V the linearity is found over a large temperature range down to 10~K, and the linear fit returns $^{51}A_{\rm hf} \simeq -0.016$~T/$\mu_{\rm B}$. To estimate the exchange coupling, $^{23}K(T)$ above 2.5~K was fitted by Eq.~\eqref{Eq9} taking $\chi_{\rm spin}$ for the 1D $S=1/2$ Heisenberg chain [Eq.~\eqref{Eq3}]. The fit returns $J/k_{\rm B}\simeq 4.22$~K and $^{23}A_{\rm hf} \simeq 0.0194$~T/$\mu_{\rm B}$. The value of $g$ was fixed to $g = 2.17$ during the fitting procedure. This value of $J/k_{\rm B}$ is close to the one obtained from the $\chi(T)$ analysis, whereas $^{23}A_{\rm hf}$ is also in good agreement with the value obtained from the $K$ vs $\chi$ analysis. An anomaly at $\sim 0.3$~K in $^{\rm 23}K(T)$ could be due to a magnetic transition.

\begin{figure}
	\includegraphics {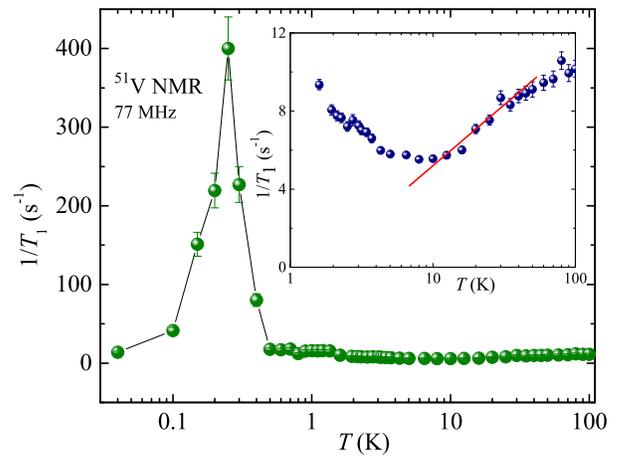}
	\caption{\label{Fig9} $1/T_{1}$ as a function of temperature measured on the $^{51}$V nuclei down to 0.044~K. Inset: $1/T_1$ above 2~K is shown in order to highlight the features around 10~K.}
\end{figure}
To study the spin dynamics, spin-lattice relaxation rate ($1/T_{1}$) was measured by irradiating the central position of the $^{51}$V spectra corresponding to the $1/2 \longleftrightarrow -1/2$ transition, choosing an appropriate pulse width. The recovery of the longitudinal magnetization was fitted by the following exponential function relevant for a quadrupole ($I = 7/2$) nuclei~\cite{Gordon783,Simmons1168}
\begin{eqnarray}
1-\frac{M(t)}{M(\infty)} &=& 0.0119\times e^{(-t/T_{1})}+ 0.068\times e^{(-6t/T_{1})} \nonumber\\
&+& 0.21 \times e^{(-15t/T_{1})}+0.71\times e^{(-28t/T_{1})}
\label{Eq10}
\end{eqnarray}
Here, $M(t)$ and $M(\infty)$ are the nuclear magnetizations at a time $t$ and $t \longrightarrow \infty$, respectively, after the saturation pulse. Temperature dependence of $^{51}$V $1/T_1$ obtained from the above fit is shown in Fig.~\ref{Fig9}. Our measurements were carried out down to 0.04~K. At high temperatures, $1/T_1$ is almost temperature-independent as expected in the paramagnetic regime~\cite{Moriya23}. At low temperatures, it exhibits a sharp peak at $T \simeq 0.25$~K due to slowing down of the fluctuating moments and is a direct evidence of the onset of magnetic LRO. In order to highlight the behavior in the intermediate temperature range, $1/T_1$ above 2~K is magnified in the inset of Fig.~\ref{Fig9}. As the temperature is lowered, $1/T_1$ decreases linearly below about 25~K, remains almost temperature-independent for 4~K$\leq T \leq 10$~K, and then starts  increasing for $T \leq 4$~K. This increase below 4~K can be attributed to the growth of AFM correlations as the system approaches the magnetic LRO state.

Further, $1/T_1T$ is directly proportional to the imaginary part of the
dynamic susceptibility $\chi_{M}(\vec{q},\omega _0)$ at the nuclear Larmor frequency $\omega_0$, which is $q$-dependent~\cite{Moriya23}. In low-dimensional spin systems, temperature-dependent $1/T_1$ often reflects dominant contributions from different $q$ values in different temperature regimes. For instance, for spin-$1/2$ Heisenberg AFM spin chains, it is theoretically predicted that with the dominant staggered contribution ($q=\pm \pi/a$) the spin-lattice relaxation rate behaves as $1/T_1 \sim T^{0}$, while the dominant contribution of the uniform component ($q=0$) results in $1/T_1 \sim T$~\cite{SandvikR9831,Sachdev13006}. The dominant contributions of $q=\pm \pi/a$ and $q=0$ are typically observed in the low-temperature ($T < J$) and high-temperature ($T \sim J$) regimes, respectively~\cite{Nath174436,Nath134451}. Thus, our experimentally observed constant and linear behaviors of $1/T_1$ with temperature over 4~K~$\leq T \leq 10$~K and 10~K$\leq T \leq 25$~K, respectively (inset of Fig.~\ref{Fig9}), are compatible with the 1D physics.

In real spin-chain systems, the non-vanishing interchain couplings often leads to the onset of magnetic LRO at very low temperatures. The interchain coupling can be calculated using the expression proposed by Schulz~\cite{Schulz2790}
\begin{equation}
	|J_{\perp}| \simeq \frac{T_{\rm N}}{1.28\sqrt{\ln(5.8J/T_{\rm N})}},
	\label{Eq11}
\end{equation}
where $J_{\perp}$ is an effective interchain coupling. Taking $T_{\rm N} \simeq 0.25$~K and $J/k_{\rm B} \simeq 5.6$~K, we arrive at the possible value of $J_{\perp}/k_{\rm B} \simeq 0.1$~K, which is indeed consistent with the value estimated from the band-structure calculations, as discussed in the following.


\subsection{Microscopic magnetic model}
\label{sec:model}
\begin{table}
	\caption{\label{tab:exchange}
		Exchange parameters of BaNa$_2$Cu(VO$_4)_2$ obtained from DFT calculations: Cu--Cu distances $d$ (in\,\r A), electron hoppings $t_i$ (in\,meV), AFM contributions to the exchange $J_i^{\rm AFM}=4t_i^2/U_{\rm eff}$ (in\,K), and total exchange couplings $J_i$ (in\,K) from the DFT+$U$ mapping procedure.
	}
	\begin{ruledtabular}
		\begin{tabular}{cc@{\hspace{4em}}cc@{\hspace{4em}}r}
			& $d_{\rm Cu-Cu}$ & $t_i$ & $J_i^{\rm AFM}$ & $J_i$ \\\hline
			$J$      & 5.507 & $-40$ &     14.9        &  6.8  \\
			$J_{ab}$ & 5.507 & $-5$  &     0.2         & $<\!0.2$  \\
			$J_{ab}'$ & 5.686 & $-1$ &     0.01        & $<\!0.2$  \\
			$J_c$    & 7.024 &  3    &     0.08        & $<\!0.2$  \\
		\end{tabular}
	\end{ruledtabular}
\end{table}

\begin{figure}
	\includegraphics{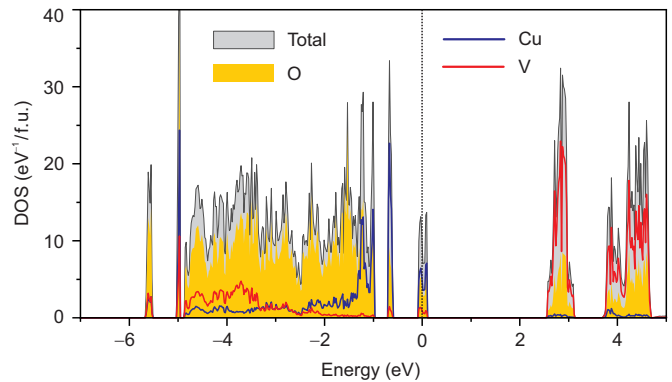}
	\caption{\label{Fig10}
		LDA density of states for BaNa$_2$Cu(VO$_4)_2$. Note the very narrow Cu $d_{x^2-y^2}$ band around 0\,eV (Fermi level) that indicates small electron hoppings and correspondingly weak exchange couplings.}
\end{figure}

LDA band structure of BaNa$_2$Cu(VO$_4)_2$ (Fig.~\ref{Fig10}) features Cu $3d$ states below the Fermi level and V $3d$ states above 2\,eV, confirming the non-magnetic state of vanadium. The overall energy spectrum is metallic, as typical for a transition-metal compound when correlation effects in the $3d$ shell were not taken into account. Nevertheless, this band structure gives an overview of possible exchange interactions, as the hopping parameters $t_i$ are proportional to the LDA bandwidth, whereas $J_i^{\rm AFM}=4t_i^2/U_{\rm eff}$. The Fermi level is crossed by two narrow bands formed by the half-filled $d_{x^2-y^2}$ orbitals of Cu$^{2+}$. The width of these bands is less than 0.2\,eV, one of the smallest in cuprates, and indicates very weak exchange couplings in BaNa$_2$Cu(VO$_4)_2$. 

DFT results for the exchange couplings are summarized in Table~\ref{tab:exchange}. Only one sizable coupling, $J/k_{\rm B} \simeq 6.8$\,K is found. It corresponds to spin chains running along $[110]$ in one layer and along $[1\bar 10]$ in the adjacent layer, the direction being chosen by the position of the double VO$_4$ bridges that connect the CuO$_4$ plaquette units (Fig.~\ref{Fig1}). Such a coupling mechanism is fairly common among the Cu$^{2+}$ compounds and can give rise to both FM and AFM superexchange depending on the orientation of the VO$_4$ tetrahedra relative to the CuO$_4$ planes~\cite{Tsirlin014401}. Larger rotations of the tetrahedra favor FM couplings. 

In BaNa$_2$Cu(VO$_4)_2$, we find $\varphi=99.0^{\circ}$, which is similar to $\varphi^{(2)}=102.2^{\circ}$ for the AFM coupling $J_a^{(2)}/k_{\rm B}=9.5$\,K in BaAg$_2$Cu(VO$_4)_2$ and very different from $\varphi^{(1)}=123.7^{\circ}$ for the FM coupling $J_a^{(1)}/k_{\rm B}=-19$\,K in the same compound~\cite{Tsirlin014401}. Here, $\varphi$ is the angle between the face of the VO$_4$ tetrahedron and the plane connecting the adjacent CuO$_4$ plaquettes, as shown in Fig.~\ref{Fig1}. Compared to BaAg$_2$Cu(VO$_4)_2$, the AFM coupling weakens from 9.5\,K to $\sim 6$\,K, likely because of the longer Cu--Cu distance (5.507\,\r A vs. 5.448\,\r A) and the increased lateral displacement $r$ of the CuO$_4$ plaquettes (0.895\,\r A vs. 0.860\,\r A).

All couplings beyond the aforementioned spin chains appear to be very weak, below 0.2\,K, and unfeasible for the DFT+$U$ mapping analysis. Their relative strengths can be assessed from the hopping parameters that suggest the dominant interchain couplings $J_{ab}$ in the $ab$ plane (along $[1\bar 10]$ for the spin chains along $[110]$, and vice versa) and $J_c$ along the $c$ direction. The in-plane coupling $J_{ab}'$ is negligible. The two stronger interchain couplings, $J_{ab}$ and $J_c$, form a non-frustrated 3D network. From $4t_i^2/U_{\rm eff}$ with $U_{\rm eff}=5$\,eV~\cite{Lebernegg224406,Ahmed214413}, one expects the coupling strength of 0.2\,K or lower, in agreement with the DFT+$U$ results. Altogether, our modeling results establish weak and non-frustrated interchain couplings in BaNa$_2$Cu(VO$_4)_2$, with $J_{\perp}/J \simeq 0.02$. The average interchain coupling of $J_{\perp}/k_{\rm B} \simeq 0.1$\,K leads to $T_N/J\simeq 0.22$\,K~\cite{Yasuda217201} in good agreement with 0.25\,K found experimentally. Therefore, we argue that long-range magnetic order in BaNa$_2$Cu(VO$_4)_2$ should be driven by weak interchain couplings, and the N\'eel temperature $T_{\rm N}/J$ is determined by the $J_{\perp}/J$ ratio.

\begin{figure}
	\includegraphics{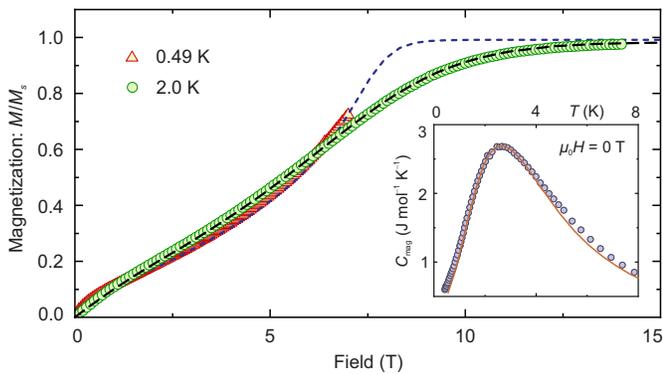}
	\caption{\label{Fig11}
		Magnetization normalized to the saturation value (main figure) and magnetic specific heat (inset) of BaNa$_2$Cu(VO$_4)_2$. Predictions of the spin-chain model with $J/k_{\rm B}=5.5$\,K and $g=2.17$ are shown with lines. In  magnetization curves, an additional 5\,\% paramagnetic contribution described by the Brillouin function was included in order to reproduce the weak bend in low magnetic fields.}
\end{figure}

Above $T_{\rm N}$, a purely one-dimensional description should hold. Indeed, we were able to fit magnetization curves down to 0.49\,K using the spin-chain model with $J/k_{\rm B}=5.5$\,K and $g=2.17$ in excellent agreement with 5.6\,K from the fit to the magnetic susceptibility and $g=2.17$ from the ESR experiment (Fig.~\ref{Fig11}). This confirms that the interchain couplings are very weak and play only marginal role even at $T<J$. Magnetic specific heat is also well described by the spin-chain model showing small deviations below 1\,K only. These deviations correspond to the upturn in $C_{\rm mag}/T$ upon approaching $T_{\rm N}$ (Fig.~\ref{Fig6}).


\section{Conclusions}
We have shown that BaNa$_2$Cu(VO$_4)_2$ strongly deviates from all of its structural siblings in terms of the magnetic behavior. The majority of these compounds are triangular magnets, while the only Cu$^{2+}$ member studied to date, BaAg$_2$Cu(VO$_4)_2$, revealed a very unusual coexistence of different spin chains, one ferromagnetic and one antiferromagnetic~\cite{Tsirlin014401,Krupskaya759}. Our present results for BaNa$_2$Cu(VO$_4)_2$ corroborate non-trivial magnetostructural correlations in Cu$^{2+}$ vanadates, where the sign of a magnetic coupling strongly depends on the spatial orientation of the VO$_4$ tetrahedra relative to the spin chains and CuO$_4$ plaquette units.

The disparity of spin chains is absent in BaNa$_2$Cu(VO$_4)_2$, but now the chains adopt two different directions and form an unusual crossed pattern. Interestingly, this crossed pattern does not cause any magnetic frustration, because the Cu$^{2+}$ ion of one chain sits exactly on top of the Cu$^{2+}$ ion of the adjacent chain (Fig.~\ref{Fig1}). Then, each magnetic site has only one coupling to a spin chain of another direction, and not two couplings, as expected theoretically~\cite{Starykh167203}. This fact highlights the importance of lateral displacements between the Cu$^{2+}$ ions of the crossed chains to induce the frustration. Such displacements do not occur in BaNa$_2$Cu(VO$_4)_2$, but they may potentially appear in sister compounds, because even the substitution of Na$^+$ by Ag$^+$ causes significant structural changes, although the two ions are very similar in size. Alternatively, one may consider structure types with a weaker spatial separation between the crossed chains that, in turn, allows several non-equivalent interactions to form a frustrated topology even in the absence of lateral displacements~\cite{tsirlin2011,mukharjee2019,weickert2019}.


\acknowledgements
We would like to acknowledge SERB, India, for financial support bearing sanction Grant No. CRG/2019/000960. Work at the Ames Laboratory was supported by the U.S. Department of Energy, Office of Science, Basic Energy Sciences, Materials Sciences and Engineering Division. The Ames Laboratory is operated for the U.S. Department of Energy by Iowa State University under Contract No. DEAC02-07CH11358. AT was funded by the Federal Ministry for Education and Research through the Sofja Kovalevskaya Award of Alexander von Humboldt Foundation.


\begin{thebibliography}{66}%
	\makeatletter
	\providecommand \@ifxundefined [1]{%
		\@ifx{#1\undefined}
	}%
	\providecommand \@ifnum [1]{%
		\ifnum #1\expandafter \@firstoftwo
		\else \expandafter \@secondoftwo
		\fi
	}%
	\providecommand \@ifx [1]{%
		\ifx #1\expandafter \@firstoftwo
		\else \expandafter \@secondoftwo
		\fi
	}%
	\providecommand \natexlab [1]{#1}%
	\providecommand \enquote  [1]{``#1''}%
	\providecommand \bibnamefont  [1]{#1}%
	\providecommand \bibfnamefont [1]{#1}%
	\providecommand \citenamefont [1]{#1}%
	\providecommand \href@noop [0]{\@secondoftwo}%
	\providecommand \href [0]{\begingroup \@sanitize@url \@href}%
	\providecommand \@href[1]{\@@startlink{#1}\@@href}%
	\providecommand \@@href[1]{\endgroup#1\@@endlink}%
	\providecommand \@sanitize@url [0]{\catcode `\\12\catcode `\$12\catcode
		`\&12\catcode `\#12\catcode `\^12\catcode `\_12\catcode `\%12\relax}%
	\providecommand \@@startlink[1]{}%
	\providecommand \@@endlink[0]{}%
	\providecommand \url  [0]{\begingroup\@sanitize@url \@url }%
	\providecommand \@url [1]{\endgroup\@href {#1}{\urlprefix }}%
	\providecommand \urlprefix  [0]{URL }%
	\providecommand \Eprint [0]{\href }%
	\providecommand \doibase [0]{http://dx.doi.org/}%
	\providecommand \selectlanguage [0]{\@gobble}%
	\providecommand \bibinfo  [0]{\@secondoftwo}%
	\providecommand \bibfield  [0]{\@secondoftwo}%
	\providecommand \translation [1]{[#1]}%
	\providecommand \BibitemOpen [0]{}%
	\providecommand \bibitemStop [0]{}%
	\providecommand \bibitemNoStop [0]{.\EOS\space}%
	\providecommand \EOS [0]{\spacefactor3000\relax}%
	\providecommand \BibitemShut  [1]{\csname bibitem#1\endcsname}%
	\let\auto@bib@innerbib\@empty
	\bibitem [{\citenamefont {{S. Sachdev}}({2007})}]{Sachdev2007}%
	\BibitemOpen
	\bibfield  {author} {\bibinfo {author} {\bibnamefont {{S. Sachdev}}},\
	}\href@noop {} {\emph {\bibinfo {title} {{Quantum phase transitions}}}}\
	(\bibinfo  {publisher} {{Wiley Online Library}},\ \bibinfo {year}
	{{2007}})\BibitemShut {NoStop}%
	\bibitem [{\citenamefont {{Ramirez}}(1994)}]{Ramirez453}%
	\BibitemOpen
	\bibfield  {author} {\bibinfo {author} {\bibfnamefont {A.~P.}\ \bibnamefont
			{{Ramirez}}},\ }\bibfield  {title} {\enquote {\bibinfo {title} {{Strongly
					Geometrically Frustrated Magnets}},}\ }\href {\doibase
		10.1146/annurev.ms.24.080194.002321} {\bibfield  {journal} {\bibinfo
			{journal} {Annu. Rev. Mater. Sci.}\ }\textbf {\bibinfo {volume} {24}},\
		\bibinfo {pages} {453} (\bibinfo {year} {1994})}\BibitemShut {NoStop}%
	\bibitem [{\citenamefont {Mermin}\ and\ \citenamefont
		{Wagner}(1966)}]{Mermin1133}%
	\BibitemOpen
	\bibfield  {author} {\bibinfo {author} {\bibfnamefont {N.~D.}\ \bibnamefont
			{Mermin}}\ and\ \bibinfo {author} {\bibfnamefont {H.}~\bibnamefont
			{Wagner}},\ }\bibfield  {title} {\enquote {\bibinfo {title} {Absence of
				ferromagnetism or antiferromagnetism in one- or two-dimensional isotropic
				{Heisenberg} models},}\ }\href {\doibase 10.1103/PhysRevLett.17.1133}
	{\bibfield  {journal} {\bibinfo  {journal} {Phys. Rev. Lett.}\ }\textbf
		{\bibinfo {volume} {17}},\ \bibinfo {pages} {1133--1136} (\bibinfo {year}
		{1966})}\BibitemShut {NoStop}%
	\bibitem [{\citenamefont {Yasuda}\ \emph {et~al.}(2005)\citenamefont {Yasuda},
		\citenamefont {Todo}, \citenamefont {Hukushima}, \citenamefont {Alet},
		\citenamefont {Keller}, \citenamefont {Troyer},\ and\ \citenamefont
		{Takayama}}]{Yasuda217201}%
	\BibitemOpen
	\bibfield  {author} {\bibinfo {author} {\bibfnamefont {C.}~\bibnamefont
			{Yasuda}}, \bibinfo {author} {\bibfnamefont {S.}~\bibnamefont {Todo}},
		\bibinfo {author} {\bibfnamefont {K.}~\bibnamefont {Hukushima}}, \bibinfo
		{author} {\bibfnamefont {F.}~\bibnamefont {Alet}}, \bibinfo {author}
		{\bibfnamefont {M.}~\bibnamefont {Keller}}, \bibinfo {author} {\bibfnamefont
			{M.}~\bibnamefont {Troyer}}, \ and\ \bibinfo {author} {\bibfnamefont
			{H.}~\bibnamefont {Takayama}},\ }\bibfield  {title} {\enquote {\bibinfo
			{title} {{N\'eel Temperature of Quasi-Low-Dimensional Heisenberg
					Antiferromagnets}},}\ }\href {\doibase 10.1103/PhysRevLett.94.217201}
	{\bibfield  {journal} {\bibinfo  {journal} {Phys. Rev. Lett.}\ }\textbf
		{\bibinfo {volume} {94}},\ \bibinfo {pages} {217201} (\bibinfo {year}
		{2005})}\BibitemShut {NoStop}%
	\bibitem [{\citenamefont {Schulz}(1996)}]{Schulz2790}%
	\BibitemOpen
	\bibfield  {author} {\bibinfo {author} {\bibfnamefont {H.~J.}\ \bibnamefont
			{Schulz}},\ }\bibfield  {title} {\enquote {\bibinfo {title} {{Dynamics of
					Coupled Quantum Spin Chains}},}\ }\href {\doibase
		10.1103/PhysRevLett.77.2790} {\bibfield  {journal} {\bibinfo  {journal}
			{Phys. Rev. Lett.}\ }\textbf {\bibinfo {volume} {77}},\ \bibinfo {pages}
		{2790} (\bibinfo {year} {1996})}\BibitemShut {NoStop}%
	\bibitem [{\citenamefont {{Greedan, John E.}}(2001)}]{Greedan37}%
	\BibitemOpen
	\bibfield  {author} {\bibinfo {author} {\bibnamefont {{Greedan, John E.}}},\
	}\bibfield  {title} {\enquote {\bibinfo {title} {{Geometrically frustrated
					magnetic materials}},}\ }\href {\doibase {10.1039/B003682J}} {\bibfield
		{journal} {\bibinfo  {journal} {{J. Mater. Chem.}}\ }\textbf {\bibinfo
			{volume} {11}},\ \bibinfo {pages} {37} (\bibinfo {year} {2001})}\BibitemShut
	{NoStop}%
	\bibitem [{\citenamefont {Kojima}\ \emph {et~al.}(1997)\citenamefont {Kojima},
		\citenamefont {Fudamoto}, \citenamefont {Larkin}, \citenamefont {Luke},
		\citenamefont {Merrin}, \citenamefont {Nachumi}, \citenamefont {Uemura},
		\citenamefont {Motoyama}, \citenamefont {Eisaki}, \citenamefont {Uchida},
		\citenamefont {Yamada}, \citenamefont {Endoh}, \citenamefont {Hosoya},
		\citenamefont {Sternlieb},\ and\ \citenamefont {Shirane}}]{Kojima1787}%
	\BibitemOpen
	\bibfield  {author} {\bibinfo {author} {\bibfnamefont {K.~M.}\ \bibnamefont
			{Kojima}}, \bibinfo {author} {\bibfnamefont {Y.}~\bibnamefont {Fudamoto}},
		\bibinfo {author} {\bibfnamefont {M.}~\bibnamefont {Larkin}}, \bibinfo
		{author} {\bibfnamefont {G.~M.}\ \bibnamefont {Luke}}, \bibinfo {author}
		{\bibfnamefont {J.}~\bibnamefont {Merrin}}, \bibinfo {author} {\bibfnamefont
			{B.}~\bibnamefont {Nachumi}}, \bibinfo {author} {\bibfnamefont {Y.~J.}\
			\bibnamefont {Uemura}}, \bibinfo {author} {\bibfnamefont {N.}~\bibnamefont
			{Motoyama}}, \bibinfo {author} {\bibfnamefont {H.}~\bibnamefont {Eisaki}},
		\bibinfo {author} {\bibfnamefont {S.}~\bibnamefont {Uchida}}, \bibinfo
		{author} {\bibfnamefont {K.}~\bibnamefont {Yamada}}, \bibinfo {author}
		{\bibfnamefont {Y.}~\bibnamefont {Endoh}}, \bibinfo {author} {\bibfnamefont
			{S.}~\bibnamefont {Hosoya}}, \bibinfo {author} {\bibfnamefont {B.~J.}\
			\bibnamefont {Sternlieb}}, \ and\ \bibinfo {author} {\bibfnamefont
			{G.}~\bibnamefont {Shirane}},\ }\bibfield  {title} {\enquote {\bibinfo
			{title} {{Reduction of Ordered Moment and N\'eel Temperature of
					Quasi-One-Dimensional Antiferromagnets ${\mathrm{Sr}}_{2}{\mathrm{CuO}}_{3}$
					and ${\mathrm{Ca}}_{2}{\mathrm{CuO}}_{3}$}},}\ }\href {\doibase
		10.1103/PhysRevLett.78.1787} {\bibfield  {journal} {\bibinfo  {journal}
			{Phys. Rev. Lett.}\ }\textbf {\bibinfo {volume} {78}},\ \bibinfo {pages}
		{1787} (\bibinfo {year} {1997})}\BibitemShut {NoStop}%
	\bibitem [{\citenamefont {Lancaster}\ \emph {et~al.}(2006)\citenamefont
		{Lancaster}, \citenamefont {Blundell}, \citenamefont {Brooks}, \citenamefont
		{Baker}, \citenamefont {Pratt}, \citenamefont {Manson}, \citenamefont
		{Landee},\ and\ \citenamefont {Baines}}]{Lancaster020410}%
	\BibitemOpen
	\bibfield  {author} {\bibinfo {author} {\bibfnamefont {T.}~\bibnamefont
			{Lancaster}}, \bibinfo {author} {\bibfnamefont {S.~J.}\ \bibnamefont
			{Blundell}}, \bibinfo {author} {\bibfnamefont {M.~L.}\ \bibnamefont
			{Brooks}}, \bibinfo {author} {\bibfnamefont {P.~J.}\ \bibnamefont {Baker}},
		\bibinfo {author} {\bibfnamefont {F.~L.}\ \bibnamefont {Pratt}}, \bibinfo
		{author} {\bibfnamefont {J.~L.}\ \bibnamefont {Manson}}, \bibinfo {author}
		{\bibfnamefont {C.~P.}\ \bibnamefont {Landee}}, \ and\ \bibinfo {author}
		{\bibfnamefont {C.}~\bibnamefont {Baines}},\ }\bibfield  {title} {\enquote
		{\bibinfo {title} {{Magnetic order in the quasi-one-dimensional
					spin-$\frac{1}{2}$ molecular chain compound copper pyrazine dinitrate}},}\
	}\href {\doibase 10.1103/PhysRevB.73.020410} {\bibfield  {journal} {\bibinfo
			{journal} {Phys. Rev. B}\ }\textbf {\bibinfo {volume} {73}},\ \bibinfo
		{pages} {020410} (\bibinfo {year} {2006})}\BibitemShut {NoStop}%
	\bibitem [{\citenamefont {Furukawa}\ \emph {et~al.}(2010)\citenamefont
		{Furukawa}, \citenamefont {Sato},\ and\ \citenamefont
		{Onoda}}]{Furukawa257205}%
	\BibitemOpen
	\bibfield  {author} {\bibinfo {author} {\bibfnamefont {S.}~\bibnamefont
			{Furukawa}}, \bibinfo {author} {\bibfnamefont {M.}~\bibnamefont {Sato}}, \
		and\ \bibinfo {author} {\bibfnamefont {S.}~\bibnamefont {Onoda}},\ }\bibfield
	{title} {\enquote {\bibinfo {title} {Chiral order and electromagnetic
				dynamics in one-dimensional multiferroic cuprates},}\ }\href {\doibase
		10.1103/PhysRevLett.105.257205} {\bibfield  {journal} {\bibinfo  {journal}
			{Phys. Rev. Lett.}\ }\textbf {\bibinfo {volume} {105}},\ \bibinfo {pages}
		{257205} (\bibinfo {year} {2010})}\BibitemShut {NoStop}%
	\bibitem [{\citenamefont {Hase}\ \emph {et~al.}(1993)\citenamefont {Hase},
		\citenamefont {Terasaki},\ and\ \citenamefont {Uchinokura}}]{Hase3651}%
	\BibitemOpen
	\bibfield  {author} {\bibinfo {author} {\bibfnamefont {M.}~\bibnamefont
			{Hase}}, \bibinfo {author} {\bibfnamefont {I.}~\bibnamefont {Terasaki}}, \
		and\ \bibinfo {author} {\bibfnamefont {K.}~\bibnamefont {Uchinokura}},\
	}\bibfield  {title} {\enquote {\bibinfo {title} {{Observation of the
					spin-Peierls transition in linear ${\mathrm{Cu}}^{2+}$ (spin-$\frac{1}{2}$)
					chains in an inorganic compound ${\mathrm{CuGeO}}_{3}$}},}\ }\href {\doibase
		10.1103/PhysRevLett.70.3651} {\bibfield  {journal} {\bibinfo  {journal}
			{Phys. Rev. Lett.}\ }\textbf {\bibinfo {volume} {70}},\ \bibinfo {pages}
		{3651} (\bibinfo {year} {1993})}\BibitemShut {NoStop}%
	\bibitem [{\citenamefont {Drechsler}\ \emph {et~al.}(2007)\citenamefont
		{Drechsler}, \citenamefont {Volkova}, \citenamefont {Vasiliev}, \citenamefont
		{Tristan}, \citenamefont {Richter}, \citenamefont {Schmitt}, \citenamefont
		{Rosner}, \citenamefont {M\'alek}, \citenamefont {Klingeler}, \citenamefont
		{Zvyagin},\ and\ \citenamefont {B\"uchner}}]{Drechsler077202}%
	\BibitemOpen
	\bibfield  {author} {\bibinfo {author} {\bibfnamefont {S.-L.}\ \bibnamefont
			{Drechsler}}, \bibinfo {author} {\bibfnamefont {O.}~\bibnamefont {Volkova}},
		\bibinfo {author} {\bibfnamefont {A.~N.}\ \bibnamefont {Vasiliev}}, \bibinfo
		{author} {\bibfnamefont {N.}~\bibnamefont {Tristan}}, \bibinfo {author}
		{\bibfnamefont {J.}~\bibnamefont {Richter}}, \bibinfo {author} {\bibfnamefont
			{M.}~\bibnamefont {Schmitt}}, \bibinfo {author} {\bibfnamefont
			{H.}~\bibnamefont {Rosner}}, \bibinfo {author} {\bibfnamefont
			{J.}~\bibnamefont {M\'alek}}, \bibinfo {author} {\bibfnamefont
			{R.}~\bibnamefont {Klingeler}}, \bibinfo {author} {\bibfnamefont {A.~A.}\
			\bibnamefont {Zvyagin}}, \ and\ \bibinfo {author} {\bibfnamefont
			{B.}~\bibnamefont {B\"uchner}},\ }\bibfield  {title} {\enquote {\bibinfo
			{title} {{Frustrated Cuprate Route from Antiferromagnetic to Ferromagnetic
					Spin-$\frac{1}{2}$ Heisenberg Chains: ${\mathrm{Li}}_{2}{\mathrm{ZrCuO}}_{4}$
					as a Missing Link near the Quantum Critical Point}},}\ }\href {\doibase
		10.1103/PhysRevLett.98.077202} {\bibfield  {journal} {\bibinfo  {journal}
			{Phys. Rev. Lett.}\ }\textbf {\bibinfo {volume} {98}},\ \bibinfo {pages}
		{077202} (\bibinfo {year} {2007})}\BibitemShut {NoStop}%
	\bibitem [{\citenamefont {Amuneke}\ \emph {et~al.}(2011)\citenamefont
		{Amuneke}, \citenamefont {Gheorghe}, \citenamefont {Lorenz},\ and\
		\citenamefont {M{\"o}ller}}]{Amuneke2207}%
	\BibitemOpen
	\bibfield  {author} {\bibinfo {author} {\bibfnamefont {N.~E.}\ \bibnamefont
			{Amuneke}}, \bibinfo {author} {\bibfnamefont {D.~E.}\ \bibnamefont
			{Gheorghe}}, \bibinfo {author} {\bibfnamefont {B.}~\bibnamefont {Lorenz}}, \
		and\ \bibinfo {author} {\bibfnamefont {A.}~\bibnamefont {M{\"o}ller}},\
	}\bibfield  {title} {\enquote {\bibinfo {title} {{Synthesis, Crystal
					Structure, and Physical Properties of BaAg$_2$Cu[VO$_4$]$_2$: A New Member of
					the S = $1/2$ Triangular Lattice}},}\ }\href {\doibase 10.1021/ic1018554}
	{\bibfield  {journal} {\bibinfo  {journal} {Inorg. Chem.}\ }\textbf {\bibinfo
			{volume} {50}},\ \bibinfo {pages} {2207} (\bibinfo {year}
		{2011})}\BibitemShut {NoStop}%
	\bibitem [{\citenamefont {M\"oller}\ \emph {et~al.}(2012)\citenamefont
		{M\"oller}, \citenamefont {Amuneke}, \citenamefont {Daniel}, \citenamefont
		{Lorenz}, \citenamefont {de~la Cruz}, \citenamefont {Gooch},\ and\
		\citenamefont {Chu}}]{Moller214422}%
	\BibitemOpen
	\bibfield  {author} {\bibinfo {author} {\bibfnamefont {A.}~\bibnamefont
			{M\"oller}}, \bibinfo {author} {\bibfnamefont {N.~E.}\ \bibnamefont
			{Amuneke}}, \bibinfo {author} {\bibfnamefont {P.}~\bibnamefont {Daniel}},
		\bibinfo {author} {\bibfnamefont {B.}~\bibnamefont {Lorenz}}, \bibinfo
		{author} {\bibfnamefont {C.~R.}\ \bibnamefont {de~la Cruz}}, \bibinfo
		{author} {\bibfnamefont {M.}~\bibnamefont {Gooch}}, \ and\ \bibinfo {author}
		{\bibfnamefont {P.~C.~W.}\ \bibnamefont {Chu}},\ }\bibfield  {title}
	{\enquote {\bibinfo {title} {{$A$Ag${}_{2}M$[VO${}_{4}{]}_{2}$
					($A=\mathrm{Ba},\mathrm{Sr}$; $M=\mathrm{Co},\mathrm{Ni}$): A series of
					ferromagnetic insulators}},}\ }\href {\doibase 10.1103/PhysRevB.85.214422}
	{\bibfield  {journal} {\bibinfo  {journal} {Phys. Rev. B}\ }\textbf {\bibinfo
			{volume} {85}},\ \bibinfo {pages} {214422} (\bibinfo {year}
		{2012})}\BibitemShut {NoStop}%
	\bibitem [{\citenamefont {Nakayama}\ \emph {et~al.}(2013)\citenamefont
		{Nakayama}, \citenamefont {Hara}, \citenamefont {Sato}, \citenamefont
		{Narumi},\ and\ \citenamefont {Nojiri}}]{Nakayama116003}%
	\BibitemOpen
	\bibfield  {author} {\bibinfo {author} {\bibfnamefont {G.}~\bibnamefont
			{Nakayama}}, \bibinfo {author} {\bibfnamefont {S.}~\bibnamefont {Hara}},
		\bibinfo {author} {\bibfnamefont {H.}~\bibnamefont {Sato}}, \bibinfo {author}
		{\bibfnamefont {Y.}~\bibnamefont {Narumi}}, \ and\ \bibinfo {author}
		{\bibfnamefont {H.}~\bibnamefont {Nojiri}},\ }\bibfield  {title} {\enquote
		{\bibinfo {title} {{Synthesis and magnetic properties of a new series of
					triangular-lattice magnets, Na$_2$BaMV$_2$O$_8$(M = Ni, Co, and Mn)}},}\
	}\href {\doibase 10.1088/0953-8984/25/11/116003} {\bibfield  {journal}
		{\bibinfo  {journal} {J. Phys.: Condens. Matter}\ }\textbf {\bibinfo {volume}
			{25}},\ \bibinfo {pages} {116003} (\bibinfo {year} {2013})}\BibitemShut
	{NoStop}%
	\bibitem [{\citenamefont {Reu{\ss}}\ \emph {et~al.}(2018)\citenamefont
		{Reu{\ss}}, \citenamefont {Ksenofontov}, \citenamefont {Tapp}, \citenamefont
		{Wulferding}, \citenamefont {Lemmens}, \citenamefont {Panthöfer},\ and\
		\citenamefont {M\"oller}}]{Reub6300}%
	\BibitemOpen
	\bibfield  {author} {\bibinfo {author} {\bibfnamefont {A.}~\bibnamefont
			{Reu{\ss}}}, \bibinfo {author} {\bibfnamefont {V.}~\bibnamefont
			{Ksenofontov}}, \bibinfo {author} {\bibfnamefont {J.}~\bibnamefont {Tapp}},
		\bibinfo {author} {\bibfnamefont {D.}~\bibnamefont {Wulferding}}, \bibinfo
		{author} {\bibfnamefont {P.}~\bibnamefont {Lemmens}}, \bibinfo {author}
		{\bibfnamefont {M.}~\bibnamefont {Panthöfer}}, \ and\ \bibinfo {author}
		{\bibfnamefont {A.}~\bibnamefont {M\"oller}},\ }\bibfield  {title} {\enquote
		{\bibinfo {title} {{Screw-Type Motion and Its Impact on Cooperativity in
					BaNa$_2$Fe[VO$_4$]$_2$}},}\ }\href {\doibase 10.1021/acs.inorgchem.8b00191}
	{\bibfield  {journal} {\bibinfo  {journal} {Inorg. Chem.}\ }\textbf {\bibinfo
			{volume} {57}},\ \bibinfo {pages} {6300} (\bibinfo {year}
		{2018})}\BibitemShut {NoStop}%
	\bibitem [{\citenamefont {Sanjeewa}\ \emph {et~al.}(2019)\citenamefont
		{Sanjeewa}, \citenamefont {Garlea}, \citenamefont {McGuire}, \citenamefont
		{McMillen},\ and\ \citenamefont {Kolis}}]{Sanjeewa2813}%
	\BibitemOpen
	\bibfield  {author} {\bibinfo {author} {\bibfnamefont {L.~D.}\ \bibnamefont
			{Sanjeewa}}, \bibinfo {author} {\bibfnamefont {V.~O.}\ \bibnamefont
			{Garlea}}, \bibinfo {author} {\bibfnamefont {M.~A.}\ \bibnamefont {McGuire}},
		\bibinfo {author} {\bibfnamefont {C.~D.}\ \bibnamefont {McMillen}}, \ and\
		\bibinfo {author} {\bibfnamefont {J.~W.}\ \bibnamefont {Kolis}},\ }\bibfield
	{title} {\enquote {\bibinfo {title} {{Magnetic Ground State Crossover in a
					Series of Glaserite Systems with Triangular Magnetic Lattices}},}\ }\href
	{\doibase 10.1021/acs.inorgchem.8b03418} {\bibfield  {journal} {\bibinfo
			{journal} {Inorg. Chem.}\ }\textbf {\bibinfo {volume} {58}},\ \bibinfo
		{pages} {2813} (\bibinfo {year} {2019})}\BibitemShut {NoStop}%
	\bibitem [{\citenamefont {Amuneke}\ \emph {et~al.}(2014)\citenamefont
		{Amuneke}, \citenamefont {Tapp}, \citenamefont {{de la Cruz}},\ and\
		\citenamefont {M\"oller}}]{Amuneke5930}%
	\BibitemOpen
	\bibfield  {author} {\bibinfo {author} {\bibfnamefont {N.~E.}\ \bibnamefont
			{Amuneke}}, \bibinfo {author} {\bibfnamefont {J.}~\bibnamefont {Tapp}},
		\bibinfo {author} {\bibfnamefont {C.~R.}\ \bibnamefont {{de la Cruz}}}, \
		and\ \bibinfo {author} {\bibfnamefont {A.}~\bibnamefont {M\"oller}},\
	}\bibfield  {title} {\enquote {\bibinfo {title} {Experimental realization of
				a unique class of compounds: {XY}-antiferromagnetic triangular lattices,
				{KAg$_2$Fe[VO$_4]_2$} and {RbAg$_2$Fe[VO$_4]_2$}, with ferroelectric ground
				states},}\ }\href {\doibase 10.1021/cm5025712} {\bibfield  {journal}
		{\bibinfo  {journal} {Chem. Mater.}\ }\textbf {\bibinfo {volume} {26}},\
		\bibinfo {pages} {5930--5935} (\bibinfo {year} {2014})}\BibitemShut {NoStop}%
	\bibitem [{\citenamefont {Lee}\ \emph {et~al.}(2020)\citenamefont {Lee},
		\citenamefont {Klauer}, \citenamefont {Menten}, \citenamefont {Lee},
		\citenamefont {Yoon}, \citenamefont {Luetkens}, \citenamefont {Lemmens},
		\citenamefont {M\"oller},\ and\ \citenamefont {Choi}}]{Lee224420}%
	\BibitemOpen
	\bibfield  {author} {\bibinfo {author} {\bibfnamefont {S.}~\bibnamefont
			{Lee}}, \bibinfo {author} {\bibfnamefont {R.}~\bibnamefont {Klauer}},
		\bibinfo {author} {\bibfnamefont {J.}~\bibnamefont {Menten}}, \bibinfo
		{author} {\bibfnamefont {W.}~\bibnamefont {Lee}}, \bibinfo {author}
		{\bibfnamefont {S.}~\bibnamefont {Yoon}}, \bibinfo {author} {\bibfnamefont
			{H.}~\bibnamefont {Luetkens}}, \bibinfo {author} {\bibfnamefont
			{P.}~\bibnamefont {Lemmens}}, \bibinfo {author} {\bibfnamefont
			{A.}~\bibnamefont {M\"oller}}, \ and\ \bibinfo {author} {\bibfnamefont
			{K.-Y.}\ \bibnamefont {Choi}},\ }\bibfield  {title} {\enquote {\bibinfo
			{title} {Unconventional spin excitations in the {$S=\frac32$} triangular
				antiferromagnet {RbAg$_2$Cr[VO$_4]_2$}},}\ }\href {\doibase
		10.1103/PhysRevB.101.224420} {\bibfield  {journal} {\bibinfo  {journal}
			{Phys. Rev. B}\ }\textbf {\bibinfo {volume} {101}},\ \bibinfo {pages}
		{224420} (\bibinfo {year} {2020})}\BibitemShut {NoStop}%
	\bibitem [{\citenamefont {Tsirlin}\ \emph {et~al.}(2012)\citenamefont
		{Tsirlin}, \citenamefont {M\"oller}, \citenamefont {Lorenz}, \citenamefont
		{Skourski},\ and\ \citenamefont {Rosner}}]{Tsirlin014401}%
	\BibitemOpen
	\bibfield  {author} {\bibinfo {author} {\bibfnamefont {A.~A.}\ \bibnamefont
			{Tsirlin}}, \bibinfo {author} {\bibfnamefont {A.}~\bibnamefont {M\"oller}},
		\bibinfo {author} {\bibfnamefont {B.}~\bibnamefont {Lorenz}}, \bibinfo
		{author} {\bibfnamefont {Y.}~\bibnamefont {Skourski}}, \ and\ \bibinfo
		{author} {\bibfnamefont {H.}~\bibnamefont {Rosner}},\ }\bibfield  {title}
	{\enquote {\bibinfo {title} {Superposition of ferromagnetic and
				antiferromagnetic spin chains in the quantum magnet
				{BaAg$_2$Cu[VO$_4]_2$}},}\ }\href {\doibase 10.1103/PhysRevB.85.014401}
	{\bibfield  {journal} {\bibinfo  {journal} {Phys. Rev. B}\ }\textbf {\bibinfo
			{volume} {85}},\ \bibinfo {pages} {014401} (\bibinfo {year}
		{2012})}\BibitemShut {NoStop}%
	\bibitem [{\citenamefont {Krupskaya}\ \emph {et~al.}(2017)\citenamefont
		{Krupskaya}, \citenamefont {Sch\"apers}, \citenamefont {Wolter},
		\citenamefont {Grafe}, \citenamefont {Vavilova}, \citenamefont {M\"oller},
		\citenamefont {B\"uchner},\ and\ \citenamefont {Kataev}}]{Krupskaya759}%
	\BibitemOpen
	\bibfield  {author} {\bibinfo {author} {\bibfnamefont {Y.}~\bibnamefont
			{Krupskaya}}, \bibinfo {author} {\bibfnamefont {M.}~\bibnamefont
			{Sch\"apers}}, \bibinfo {author} {\bibfnamefont {A.U.B.}\ \bibnamefont
			{Wolter}}, \bibinfo {author} {\bibfnamefont {H.-J.}\ \bibnamefont {Grafe}},
		\bibinfo {author} {\bibfnamefont {E.}~\bibnamefont {Vavilova}}, \bibinfo
		{author} {\bibfnamefont {A.}~\bibnamefont {M\"oller}}, \bibinfo {author}
		{\bibfnamefont {B.}~\bibnamefont {B\"uchner}}, \ and\ \bibinfo {author}
		{\bibfnamefont {V.}~\bibnamefont {Kataev}},\ }\bibfield  {title} {\enquote
		{\bibinfo {title} {Magnetic resonance study of the spin-1/2 quantum magnet
				{BaAg$_2$Cu[VO$_4]_2$}},}\ }\href {\doibase 10.1515/zpch-2016-0829}
	{\bibfield  {journal} {\bibinfo  {journal} {Z. Phys. Chem.}\ }\textbf
		{\bibinfo {volume} {231}},\ \bibinfo {pages} {759} (\bibinfo {year}
		{2017})}\BibitemShut {NoStop}%
	\bibitem [{\citenamefont {von Postel}\ and\ \citenamefont
		{M\"uller-Buschbaum}(1992)}]{Von107}%
	\BibitemOpen
	\bibfield  {author} {\bibinfo {author} {\bibfnamefont {M.}~\bibnamefont {von
				Postel}}\ and\ \bibinfo {author} {\bibfnamefont {Hk.}\ \bibnamefont
			{M\"uller-Buschbaum}},\ }\bibfield  {title} {\enquote {\bibinfo {title}
			{{Na${_2}$BaCuV${_2}$O${_8}$: Ein neuer Strukturtyp der
					Alkali-Erdalkalimetall Kupfer-Oxovanadate}},}\ }\href {\doibase
		10.1002/zaac.19926180119} {\bibfield  {journal} {\bibinfo  {journal} {Z.
				Anorg. Allg. Chem.}\ }\textbf {\bibinfo {volume} {618}},\ \bibinfo {pages}
		{107} (\bibinfo {year} {1992})}\BibitemShut {NoStop}%
	\bibitem [{\citenamefont {Starykh}\ \emph {et~al.}(2002)\citenamefont
		{Starykh}, \citenamefont {Singh},\ and\ \citenamefont
		{Levine}}]{Starykh167203}%
	\BibitemOpen
	\bibfield  {author} {\bibinfo {author} {\bibfnamefont {O.~A.}\ \bibnamefont
			{Starykh}}, \bibinfo {author} {\bibfnamefont {R.~R.~P.}\ \bibnamefont
			{Singh}}, \ and\ \bibinfo {author} {\bibfnamefont {G.~C.}\ \bibnamefont
			{Levine}},\ }\bibfield  {title} {\enquote {\bibinfo {title} {Spinons in a
				crossed-chains model of a {2D} spin liquid},}\ }\href {\doibase
		10.1103/PhysRevLett.88.167203} {\bibfield  {journal} {\bibinfo  {journal}
			{Phys. Rev. Lett.}\ }\textbf {\bibinfo {volume} {88}},\ \bibinfo {pages}
		{167203} (\bibinfo {year} {2002})}\BibitemShut {NoStop}%
	\bibitem [{\citenamefont {Sindzingre}\ \emph {et~al.}(2002)\citenamefont
		{Sindzingre}, \citenamefont {Fouet},\ and\ \citenamefont
		{Lhuillier}}]{Sindzingre174424}%
	\BibitemOpen
	\bibfield  {author} {\bibinfo {author} {\bibfnamefont {P.}~\bibnamefont
			{Sindzingre}}, \bibinfo {author} {\bibfnamefont {J.-B.}\ \bibnamefont
			{Fouet}}, \ and\ \bibinfo {author} {\bibfnamefont {C.}~\bibnamefont
			{Lhuillier}},\ }\bibfield  {title} {\enquote {\bibinfo {title}
			{One-dimensional behavior and sliding {Luttinger} liquid phase in a
				frustrated spin-1/2 crossed chain model: Contribution of exact
				diagonalizations},}\ }\href {\doibase 10.1103/PhysRevB.66.174424} {\bibfield
		{journal} {\bibinfo  {journal} {Phys. Rev. B}\ }\textbf {\bibinfo {volume}
			{66}},\ \bibinfo {pages} {174424} (\bibinfo {year} {2002})}\BibitemShut
	{NoStop}%
	\bibitem [{\citenamefont {Brenig}\ and\ \citenamefont
		{Grzeschik}(2004)}]{Brenig064420}%
	\BibitemOpen
	\bibfield  {author} {\bibinfo {author} {\bibfnamefont {W.}~\bibnamefont
			{Brenig}}\ and\ \bibinfo {author} {\bibfnamefont {M.}~\bibnamefont
			{Grzeschik}},\ }\bibfield  {title} {\enquote {\bibinfo {title} {Valence-bond
				crystal phase of the crossed-chain quantum spin model},}\ }\href {\doibase
		10.1103/PhysRevB.69.064420} {\bibfield  {journal} {\bibinfo  {journal} {Phys.
				Rev. B}\ }\textbf {\bibinfo {volume} {69}},\ \bibinfo {pages} {064420}
		(\bibinfo {year} {2004})}\BibitemShut {NoStop}%
	\bibitem [{\citenamefont {Starykh}\ \emph {et~al.}(2005)\citenamefont
		{Starykh}, \citenamefont {Furusaki},\ and\ \citenamefont
		{Balents}}]{Starykh094416}%
	\BibitemOpen
	\bibfield  {author} {\bibinfo {author} {\bibfnamefont {O.~A.}\ \bibnamefont
			{Starykh}}, \bibinfo {author} {\bibfnamefont {A.}~\bibnamefont {Furusaki}}, \
		and\ \bibinfo {author} {\bibfnamefont {L.}~\bibnamefont {Balents}},\
	}\bibfield  {title} {\enquote {\bibinfo {title} {Anisotropic pyrochlores and
				the global phase diagram of the checkerboard antiferromagnet},}\ }\href
	{\doibase 10.1103/PhysRevB.72.094416} {\bibfield  {journal} {\bibinfo
			{journal} {Phys. Rev. B}\ }\textbf {\bibinfo {volume} {72}},\ \bibinfo
		{pages} {094416} (\bibinfo {year} {2005})}\BibitemShut {NoStop}%
	\bibitem [{\citenamefont {Bishop}\ \emph {et~al.}(2012)\citenamefont {Bishop},
		\citenamefont {Li}, \citenamefont {Farnell}, \citenamefont {Richter},\ and\
		\citenamefont {Campbell}}]{Bishop205122}%
	\BibitemOpen
	\bibfield  {author} {\bibinfo {author} {\bibfnamefont {R.~F.}\ \bibnamefont
			{Bishop}}, \bibinfo {author} {\bibfnamefont {P.~H.~Y.}\ \bibnamefont {Li}},
		\bibinfo {author} {\bibfnamefont {D.~J.~J.}\ \bibnamefont {Farnell}},
		\bibinfo {author} {\bibfnamefont {J.}~\bibnamefont {Richter}}, \ and\
		\bibinfo {author} {\bibfnamefont {C.~E.}\ \bibnamefont {Campbell}},\
	}\bibfield  {title} {\enquote {\bibinfo {title} {Frustrated {Heisenberg}
				antiferromagnet on the checkerboard lattice: {$J_1-J_2$} model},}\ }\href
	{\doibase 10.1103/PhysRevB.85.205122} {\bibfield  {journal} {\bibinfo
			{journal} {Phys. Rev. B}\ }\textbf {\bibinfo {volume} {85}},\ \bibinfo
		{pages} {205122} (\bibinfo {year} {2012})}\BibitemShut {NoStop}%
	\bibitem [{\citenamefont {{J. Rodríguez-Carvajal}}({1993})}]{Rodriguez55}%
	\BibitemOpen
	\bibfield  {author} {\bibinfo {author} {\bibnamefont {{J.
					Rodríguez-Carvajal}}},\ }\bibfield  {title} {\enquote {\bibinfo {title}
			{{Recent advances in magnetic structure determination by neutron powder
					diffraction}},}\ }\href {\doibase
		https://doi.org/10.1016/0921-4526(93)90108-I} {\bibfield  {journal} {\bibinfo
			{journal} {{Physica B: Condensed Matter}}\ }\textbf {\bibinfo {volume}
			{{192}}},\ \bibinfo {pages} {{55}} (\bibinfo {year} {{1993}})}\BibitemShut
	{NoStop}%
	\bibitem [{\citenamefont {Koepernik}\ and\ \citenamefont
		{Eschrig}(1999)}]{fplo}%
	\BibitemOpen
	\bibfield  {author} {\bibinfo {author} {\bibfnamefont {K.}~\bibnamefont
			{Koepernik}}\ and\ \bibinfo {author} {\bibfnamefont {H.}~\bibnamefont
			{Eschrig}},\ }\bibfield  {title} {\enquote {\bibinfo {title} {Full-potential
				nonorthogonal local-orbital minimum-basis band-structure scheme},}\ }\href
	{\doibase 10.1103/PhysRevB.59.1743} {\bibfield  {journal} {\bibinfo
			{journal} {Phys. Rev. B}\ }\textbf {\bibinfo {volume} {59}},\ \bibinfo
		{pages} {1743} (\bibinfo {year} {1999})}\BibitemShut {NoStop}%
	\bibitem [{\citenamefont {Perdew}\ and\ \citenamefont
		{Wang}(1992)}]{Perdew13244}%
	\BibitemOpen
	\bibfield  {author} {\bibinfo {author} {\bibfnamefont {J.~P.}\ \bibnamefont
			{Perdew}}\ and\ \bibinfo {author} {\bibfnamefont {Y.}~\bibnamefont {Wang}},\
	}\bibfield  {title} {\enquote {\bibinfo {title} {Accurate and simple analytic
				representation of the electron-gas correlation energy},}\ }\href {\doibase
		10.1103/PhysRevB.45.13244} {\bibfield  {journal} {\bibinfo  {journal} {Phys.
				Rev. B}\ }\textbf {\bibinfo {volume} {45}},\ \bibinfo {pages} {13244}
		(\bibinfo {year} {1992})}\BibitemShut {NoStop}%
	\bibitem [{\citenamefont {Xiang}\ \emph {et~al.}(2011)\citenamefont {Xiang},
		\citenamefont {Kan}, \citenamefont {Wei}, \citenamefont {Whangbo},\ and\
		\citenamefont {Gong}}]{Xiang224429}%
	\BibitemOpen
	\bibfield  {author} {\bibinfo {author} {\bibfnamefont {H.~J.}\ \bibnamefont
			{Xiang}}, \bibinfo {author} {\bibfnamefont {E.~J.}\ \bibnamefont {Kan}},
		\bibinfo {author} {\bibfnamefont {S.-H.}\ \bibnamefont {Wei}}, \bibinfo
		{author} {\bibfnamefont {M.-H.}\ \bibnamefont {Whangbo}}, \ and\ \bibinfo
		{author} {\bibfnamefont {X.~G.}\ \bibnamefont {Gong}},\ }\bibfield  {title}
	{\enquote {\bibinfo {title} {Predicting the spin-lattice order of frustrated
				systems from first principles},}\ }\href {\doibase
		10.1103/PhysRevB.84.224429} {\bibfield  {journal} {\bibinfo  {journal} {Phys.
				Rev. B}\ }\textbf {\bibinfo {volume} {84}},\ \bibinfo {pages} {224429}
		(\bibinfo {year} {2011})}\BibitemShut {NoStop}%
	\bibitem [{\citenamefont {Janson}\ \emph {et~al.}(2011)\citenamefont {Janson},
		\citenamefont {Tsirlin}, \citenamefont {Osipova}, \citenamefont {Berdonosov},
		\citenamefont {Olenev}, \citenamefont {Dolgikh},\ and\ \citenamefont
		{Rosner}}]{Janson144423}%
	\BibitemOpen
	\bibfield  {author} {\bibinfo {author} {\bibfnamefont {O.}~\bibnamefont
			{Janson}}, \bibinfo {author} {\bibfnamefont {A.~A.}\ \bibnamefont {Tsirlin}},
		\bibinfo {author} {\bibfnamefont {E.~S.}\ \bibnamefont {Osipova}}, \bibinfo
		{author} {\bibfnamefont {P.~S.}\ \bibnamefont {Berdonosov}}, \bibinfo
		{author} {\bibfnamefont {A.~V.}\ \bibnamefont {Olenev}}, \bibinfo {author}
		{\bibfnamefont {V.~A.}\ \bibnamefont {Dolgikh}}, \ and\ \bibinfo {author}
		{\bibfnamefont {H.}~\bibnamefont {Rosner}},\ }\bibfield  {title} {\enquote
		{\bibinfo {title} {{CaCu$_2$(SeO$_3)_2$Cl$_2$}: Spin-$\frac12$ {Heisenberg}
				chain compound with complex frustrated interchain couplings},}\ }\href
	{\doibase 10.1103/PhysRevB.83.144423} {\bibfield  {journal} {\bibinfo
			{journal} {Phys. Rev. B}\ }\textbf {\bibinfo {volume} {83}},\ \bibinfo
		{pages} {144423} (\bibinfo {year} {2011})}\BibitemShut {NoStop}%
	\bibitem [{\citenamefont {Todo}\ and\ \citenamefont {Kato}(2001)}]{loop}%
	\BibitemOpen
	\bibfield  {author} {\bibinfo {author} {\bibfnamefont {S.}~\bibnamefont
			{Todo}}\ and\ \bibinfo {author} {\bibfnamefont {K.}~\bibnamefont {Kato}},\
	}\bibfield  {title} {\enquote {\bibinfo {title} {Cluster algorithms for
				{general-$S$} quantum spin systems},}\ }\href {\doibase
		10.1103/PhysRevLett.87.047203} {\bibfield  {journal} {\bibinfo  {journal}
			{Phys. Rev. Lett.}\ }\textbf {\bibinfo {volume} {87}},\ \bibinfo {pages}
		{047203} (\bibinfo {year} {2001})}\BibitemShut {NoStop}%
	\bibitem [{\citenamefont {Alet}\ \emph {et~al.}(2005)\citenamefont {Alet},
		\citenamefont {Wessel},\ and\ \citenamefont {Troyer}}]{dirloop}%
	\BibitemOpen
	\bibfield  {author} {\bibinfo {author} {\bibfnamefont {F.}~\bibnamefont
			{Alet}}, \bibinfo {author} {\bibfnamefont {S.}~\bibnamefont {Wessel}}, \ and\
		\bibinfo {author} {\bibfnamefont {M.}~\bibnamefont {Troyer}},\ }\bibfield
	{title} {\enquote {\bibinfo {title} {Generalized directed loop method for
				quantum {Monte Carlo} simulations},}\ }\href {\doibase
		10.1103/PhysRevE.71.036706} {\bibfield  {journal} {\bibinfo  {journal} {Phys.
				Rev. E}\ }\textbf {\bibinfo {volume} {71}},\ \bibinfo {pages} {036706}
		(\bibinfo {year} {2005})}\BibitemShut {NoStop}%
	\bibitem [{\citenamefont {Albuquerque}\ \emph {et~al.}(2007)\citenamefont
		{Albuquerque}, \citenamefont {Alet}, \citenamefont {Corboz}, \citenamefont
		{Dayal}, \citenamefont {Feiguin}, \citenamefont {Fuchs}, \citenamefont
		{Gamper}, \citenamefont {Gull}, \citenamefont {G\"urtler}, \citenamefont
		{Honecker}, \citenamefont {Igarashi}, \citenamefont {K\"orner}, \citenamefont
		{Kozhevnikov}, \citenamefont {L\"auchli}, \citenamefont {Manmana},
		\citenamefont {Matsumoto}, \citenamefont {McCulloch}, \citenamefont {Michel},
		\citenamefont {Noack}, \citenamefont {Paw{\l}owski}, \citenamefont {Pollet},
		\citenamefont {Pruschke}, \citenamefont {Schollw\"ock}, \citenamefont {Todo},
		\citenamefont {Trebst}, \citenamefont {Troyer}, \citenamefont {Werner},\ and\
		\citenamefont {Wessel}}]{alps}%
	\BibitemOpen
	\bibfield  {author} {\bibinfo {author} {\bibfnamefont {A.F.}\ \bibnamefont
			{Albuquerque}}, \bibinfo {author} {\bibfnamefont {F.}~\bibnamefont {Alet}},
		\bibinfo {author} {\bibfnamefont {P.}~\bibnamefont {Corboz}}, \bibinfo
		{author} {\bibfnamefont {P.}~\bibnamefont {Dayal}}, \bibinfo {author}
		{\bibfnamefont {A.}~\bibnamefont {Feiguin}}, \bibinfo {author} {\bibfnamefont
			{S.}~\bibnamefont {Fuchs}}, \bibinfo {author} {\bibfnamefont
			{L.}~\bibnamefont {Gamper}}, \bibinfo {author} {\bibfnamefont
			{E.}~\bibnamefont {Gull}}, \bibinfo {author} {\bibfnamefont {S.}~\bibnamefont
			{G\"urtler}}, \bibinfo {author} {\bibfnamefont {A.}~\bibnamefont {Honecker}},
		\bibinfo {author} {\bibfnamefont {R.}~\bibnamefont {Igarashi}}, \bibinfo
		{author} {\bibfnamefont {M.}~\bibnamefont {K\"orner}}, \bibinfo {author}
		{\bibfnamefont {A.}~\bibnamefont {Kozhevnikov}}, \bibinfo {author}
		{\bibfnamefont {A.}~\bibnamefont {L\"auchli}}, \bibinfo {author}
		{\bibfnamefont {S.R.}\ \bibnamefont {Manmana}}, \bibinfo {author}
		{\bibfnamefont {M.}~\bibnamefont {Matsumoto}}, \bibinfo {author}
		{\bibfnamefont {I.P.}\ \bibnamefont {McCulloch}}, \bibinfo {author}
		{\bibfnamefont {F.}~\bibnamefont {Michel}}, \bibinfo {author} {\bibfnamefont
			{R.M.}\ \bibnamefont {Noack}}, \bibinfo {author} {\bibfnamefont
			{G.}~\bibnamefont {Paw{\l}owski}}, \bibinfo {author} {\bibfnamefont
			{L.}~\bibnamefont {Pollet}}, \bibinfo {author} {\bibfnamefont
			{T.}~\bibnamefont {Pruschke}}, \bibinfo {author} {\bibfnamefont
			{U.}~\bibnamefont {Schollw\"ock}}, \bibinfo {author} {\bibfnamefont
			{S.}~\bibnamefont {Todo}}, \bibinfo {author} {\bibfnamefont {S.}~\bibnamefont
			{Trebst}}, \bibinfo {author} {\bibfnamefont {M.}~\bibnamefont {Troyer}},
		\bibinfo {author} {\bibfnamefont {P.}~\bibnamefont {Werner}}, \ and\ \bibinfo
		{author} {\bibfnamefont {S.}~\bibnamefont {Wessel}},\ }\bibfield  {title}
	{\enquote {\bibinfo {title} {{The ALPS project release 1.3: Open-source
					software for strongly correlated systems}},}\ }\href {\doibase
		10.1016/j.jmmm.2006.10.304} {\bibfield  {journal} {\bibinfo  {journal} {J.
				Magn. Magn. Mater.}\ }\textbf {\bibinfo {volume} {310}},\ \bibinfo {pages}
		{1187} (\bibinfo {year} {2007})}\BibitemShut {NoStop}%
	\bibitem [{\citenamefont {Bonner}\ and\ \citenamefont
		{Fisher}(1964)}]{BonnerA640}%
	\BibitemOpen
	\bibfield  {author} {\bibinfo {author} {\bibfnamefont {J.~C.}\ \bibnamefont
			{Bonner}}\ and\ \bibinfo {author} {\bibfnamefont {M.~E.}\ \bibnamefont
			{Fisher}},\ }\bibfield  {title} {\enquote {\bibinfo {title} {{Linear Magnetic
					Chains with Anisotropic Coupling}},}\ }\href {\doibase
		10.1103/PhysRev.135.A640} {\bibfield  {journal} {\bibinfo  {journal} {Phys.
				Rev.}\ }\textbf {\bibinfo {volume} {135}},\ \bibinfo {pages} {A640} (\bibinfo
		{year} {1964})}\BibitemShut {NoStop}%
	\bibitem [{\citenamefont {Eggert}\ \emph {et~al.}(1994)\citenamefont {Eggert},
		\citenamefont {Affleck},\ and\ \citenamefont {Takahashi}}]{Eggert332}%
	\BibitemOpen
	\bibfield  {author} {\bibinfo {author} {\bibfnamefont {S.}~\bibnamefont
			{Eggert}}, \bibinfo {author} {\bibfnamefont {I.}~\bibnamefont {Affleck}}, \
		and\ \bibinfo {author} {\bibfnamefont {M.}~\bibnamefont {Takahashi}},\
	}\bibfield  {title} {\enquote {\bibinfo {title} {{Susceptibility of the spin
					$\frac{1}{2}$ Heisenberg antiferromagnetic chain}},}\ }\href {\doibase
		10.1103/PhysRevLett.73.332} {\bibfield  {journal} {\bibinfo  {journal} {Phys.
				Rev. Lett.}\ }\textbf {\bibinfo {volume} {73}},\ \bibinfo {pages} {332}
		(\bibinfo {year} {1994})}\BibitemShut {NoStop}%
	\bibitem [{\citenamefont {{P. W. Selwood}}({2013})}]{Selwood2013}%
	\BibitemOpen
	\bibfield  {author} {\bibinfo {author} {\bibnamefont {{P. W. Selwood}}},\
	}\href@noop {} {\emph {\bibinfo {title} {{Magnetochemistry}}}}\ (\bibinfo
	{publisher} {{Read Books Ltd}},\ \bibinfo {year} {{2013}})\BibitemShut
	{NoStop}%
	\bibitem [{\citenamefont {Mendelsohn}\ \emph {et~al.}(1970)\citenamefont
		{Mendelsohn}, \citenamefont {Biggs},\ and\ \citenamefont
		{Mann}}]{Mendelsohn1130}%
	\BibitemOpen
	\bibfield  {author} {\bibinfo {author} {\bibnamefont {Mendelsohn}}, \bibinfo
		{author} {\bibnamefont {Biggs}}, \ and\ \bibinfo {author} {\bibnamefont
			{Mann}},\ }\bibfield  {title} {\enquote {\bibinfo {title} {Hartree-fock
				diamagnetic susceptibilities},}\ }\href {\doibase 10.1103/PhysRevA.2.1130}
	{\bibfield  {journal} {\bibinfo  {journal} {Phys. Rev. A}\ }\textbf {\bibinfo
			{volume} {2}},\ \bibinfo {pages} {1130} (\bibinfo {year} {1970})}\BibitemShut
	{NoStop}%
	\bibitem [{\citenamefont {Motoyama}\ \emph {et~al.}(1996)\citenamefont
		{Motoyama}, \citenamefont {Eisaki},\ and\ \citenamefont
		{Uchida}}]{Motoyama3212}%
	\BibitemOpen
	\bibfield  {author} {\bibinfo {author} {\bibfnamefont {N.}~\bibnamefont
			{Motoyama}}, \bibinfo {author} {\bibfnamefont {H.}~\bibnamefont {Eisaki}}, \
		and\ \bibinfo {author} {\bibfnamefont {S.}~\bibnamefont {Uchida}},\
	}\bibfield  {title} {\enquote {\bibinfo {title} {{Magnetic Susceptibility of
					Ideal Spin $\frac{1}{2}$ Heisenberg Antiferromagnetic Chain Systems,
					${\mathrm{Sr}}_{2}{\mathrm{CuO}}_{3}$ and ${\mathrm{SrCuO}}_{2}$}},}\ }\href
	{\doibase 10.1103/PhysRevLett.76.3212} {\bibfield  {journal} {\bibinfo
			{journal} {Phys. Rev. Lett.}\ }\textbf {\bibinfo {volume} {76}},\ \bibinfo
		{pages} {3212} (\bibinfo {year} {1996})}\BibitemShut {NoStop}%
	\bibitem [{\citenamefont {Nath}\ \emph {et~al.}(2005)\citenamefont {Nath},
		\citenamefont {Mahajan}, \citenamefont {B\"uttgen}, \citenamefont {Kegler},
		\citenamefont {Loidl},\ and\ \citenamefont {Bobroff}}]{Nath174436}%
	\BibitemOpen
	\bibfield  {author} {\bibinfo {author} {\bibfnamefont {R.}~\bibnamefont
			{Nath}}, \bibinfo {author} {\bibfnamefont {A.~V.}\ \bibnamefont {Mahajan}},
		\bibinfo {author} {\bibfnamefont {N.}~\bibnamefont {B\"uttgen}}, \bibinfo
		{author} {\bibfnamefont {C.}~\bibnamefont {Kegler}}, \bibinfo {author}
		{\bibfnamefont {A.}~\bibnamefont {Loidl}}, \ and\ \bibinfo {author}
		{\bibfnamefont {J.}~\bibnamefont {Bobroff}},\ }\bibfield  {title} {\enquote
		{\bibinfo {title} {{Study of one-dimensional nature of Spin $\frac{1}{2}$
					(Sr,Ba)${_2}$Cu(PO${_4}$)${_2}$ and BaCuP${_2}$O${_7}$ via
					$^{31}{\mathrm{P}}$ NMR}},}\ }\href {\doibase 10.1103/PhysRevB.71.174436}
	{\bibfield  {journal} {\bibinfo  {journal} {Phys. Rev. B}\ }\textbf {\bibinfo
			{volume} {7}},\ \bibinfo {pages} {174436} (\bibinfo {year}
		{2005})}\BibitemShut {NoStop}%
	\bibitem [{\citenamefont {Ahmed}\ \emph {et~al.}(2015)\citenamefont {Ahmed},
		\citenamefont {Tsirlin},\ and\ \citenamefont {Nath}}]{Ahmed214413}%
	\BibitemOpen
	\bibfield  {author} {\bibinfo {author} {\bibfnamefont {N.}~\bibnamefont
			{Ahmed}}, \bibinfo {author} {\bibfnamefont {A.~A.}\ \bibnamefont {Tsirlin}},
		\ and\ \bibinfo {author} {\bibfnamefont {R.}~\bibnamefont {Nath}},\
	}\bibfield  {title} {\enquote {\bibinfo {title} {{Multiple magnetic
					transitions in the spin-$\frac{1}{2}$ chain antiferromagnet
					${{\mathrm{SrCuTe}}}_{2}{{\mathrm{O}}}_{6}$}},}\ }\href {\doibase
		10.1103/PhysRevB.91.214413} {\bibfield  {journal} {\bibinfo  {journal} {Phys.
				Rev. B}\ }\textbf {\bibinfo {volume} {91}},\ \bibinfo {pages} {214413}
		(\bibinfo {year} {2015})}\BibitemShut {NoStop}%
	\bibitem [{\citenamefont {Takigawa}\ \emph {et~al.}(1989)\citenamefont
		{Takigawa}, \citenamefont {Hammel}, \citenamefont {Heffner}, \citenamefont
		{Fisk}, \citenamefont {Smith},\ and\ \citenamefont {Schwarz}}]{Takigawa1989}%
	\BibitemOpen
	\bibfield  {author} {\bibinfo {author} {\bibfnamefont {M.}~\bibnamefont
			{Takigawa}}, \bibinfo {author} {\bibfnamefont {P.~C.}\ \bibnamefont
			{Hammel}}, \bibinfo {author} {\bibfnamefont {R.~H.}\ \bibnamefont {Heffner}},
		\bibinfo {author} {\bibfnamefont {Z.}~\bibnamefont {Fisk}}, \bibinfo {author}
		{\bibfnamefont {J.~L.}\ \bibnamefont {Smith}}, \ and\ \bibinfo {author}
		{\bibfnamefont {R.~B.}\ \bibnamefont {Schwarz}},\ }\bibfield  {title}
	{\enquote {\bibinfo {title} {Anisotropic {Cu} {Knight} shift and magnetic
				susceptibility in the normal state of {YBa$_2$Cu$_3$O$_7$}},}\ }\href
	{\doibase 10.1103/PhysRevB.39.300} {\bibfield  {journal} {\bibinfo  {journal}
			{Phys. Rev. B}\ }\textbf {\bibinfo {volume} {39}},\ \bibinfo {pages} {300}
		(\bibinfo {year} {1989})}\BibitemShut {NoStop}%
	\bibitem [{\citenamefont {Kl\"umper}(1998)}]{klumper1998}%
	\BibitemOpen
	\bibfield  {author} {\bibinfo {author} {\bibfnamefont {A.}~\bibnamefont
			{Kl\"umper}},\ }\bibfield  {title} {\enquote {\bibinfo {title} {The spin-1/2
				{Heisenberg} chain: thermodynamics, quantum criticality and {spin-Peierls}
				exponents},}\ }\href {\doibase 10.1007/s100510050491} {\bibfield  {journal}
		{\bibinfo  {journal} {Eur. Phys. J. B}\ }\textbf {\bibinfo {volume} {5}},\
		\bibinfo {pages} {677--685} (\bibinfo {year} {1998})}\BibitemShut {NoStop}%
	\bibitem [{\citenamefont {Lebernegg}\ \emph {et~al.}(2011)\citenamefont
		{Lebernegg}, \citenamefont {Tsirlin}, \citenamefont {Janson}, \citenamefont
		{Nath}, \citenamefont {Sichelschmidt}, \citenamefont {Skourski},
		\citenamefont {Amthauer},\ and\ \citenamefont {Rosner}}]{Lebernegg174436}%
	\BibitemOpen
	\bibfield  {author} {\bibinfo {author} {\bibfnamefont {S.}~\bibnamefont
			{Lebernegg}}, \bibinfo {author} {\bibfnamefont {A.~A.}\ \bibnamefont
			{Tsirlin}}, \bibinfo {author} {\bibfnamefont {O.}~\bibnamefont {Janson}},
		\bibinfo {author} {\bibfnamefont {R.}~\bibnamefont {Nath}}, \bibinfo {author}
		{\bibfnamefont {J.}~\bibnamefont {Sichelschmidt}}, \bibinfo {author}
		{\bibfnamefont {Yu.}\ \bibnamefont {Skourski}}, \bibinfo {author}
		{\bibfnamefont {G.}~\bibnamefont {Amthauer}}, \ and\ \bibinfo {author}
		{\bibfnamefont {H.}~\bibnamefont {Rosner}},\ }\bibfield  {title} {\enquote
		{\bibinfo {title} {{Magnetic model for ${A}_{2}$CuP${}_{2}$O${}_{7}$
					($A=\text{Na}$, Li): One-dimensional versus two-dimensional behavior}},}\
	}\href {\doibase 10.1103/PhysRevB.84.174436} {\bibfield  {journal} {\bibinfo
			{journal} {Phys. Rev. B}\ }\textbf {\bibinfo {volume} {84}},\ \bibinfo
		{pages} {174436} (\bibinfo {year} {2011})}\BibitemShut {NoStop}%
	\bibitem [{\citenamefont {Abragam}\ and\ \citenamefont
		{Bleaney}(2012)}]{Abragam2012electron}%
	\BibitemOpen
	\bibfield  {author} {\bibinfo {author} {\bibfnamefont {A.}~\bibnamefont
			{Abragam}}\ and\ \bibinfo {author} {\bibfnamefont {B.}~\bibnamefont
			{Bleaney}},\ }\href@noop {} {\emph {\bibinfo {title} {{Electron Paramagnetic
					Resonance of Transition Ions}}}}\ (\bibinfo  {publisher} {OUP Oxford},\
	\bibinfo {year} {2012})\BibitemShut {NoStop}%
	\bibitem [{\citenamefont {Kochelaev}\ \emph {et~al.}(1997)\citenamefont
		{Kochelaev}, \citenamefont {Sichelschmidt}, \citenamefont {Elschner},
		\citenamefont {Lemor},\ and\ \citenamefont {Loidl}}]{Kochelaev4274}%
	\BibitemOpen
	\bibfield  {author} {\bibinfo {author} {\bibfnamefont {B.~I.}\ \bibnamefont
			{Kochelaev}}, \bibinfo {author} {\bibfnamefont {J.}~\bibnamefont
			{Sichelschmidt}}, \bibinfo {author} {\bibfnamefont {B.}~\bibnamefont
			{Elschner}}, \bibinfo {author} {\bibfnamefont {W.}~\bibnamefont {Lemor}}, \
		and\ \bibinfo {author} {\bibfnamefont {A.}~\bibnamefont {Loidl}},\ }\bibfield
	{title} {\enquote {\bibinfo {title} {{Intrinsic EPR in
					${\mathrm{La}}_{2\ensuremath{-}\mathit{x}}{\mathrm{Sr}}_{\mathit{x}}{\mathrm{CuO}}_{4}$:
					Manifestation of Three-Spin Polarons}},}\ }\href {\doibase
		10.1103/PhysRevLett.79.4274} {\bibfield  {journal} {\bibinfo  {journal}
			{Phys. Rev. Lett.}\ }\textbf {\bibinfo {volume} {79}},\ \bibinfo {pages}
		{4274} (\bibinfo {year} {1997})}\BibitemShut {NoStop}%
	\bibitem [{\citenamefont {Nath}\ \emph {et~al.}(2014)\citenamefont {Nath},
		\citenamefont {Ranjith}, \citenamefont {Sichelschmidt}, \citenamefont
		{Baenitz}, \citenamefont {Skourski}, \citenamefont {Alet}, \citenamefont
		{Rousochatzakis},\ and\ \citenamefont {Tsirlin}}]{Nath014407}%
	\BibitemOpen
	\bibfield  {author} {\bibinfo {author} {\bibfnamefont {R.}~\bibnamefont
			{Nath}}, \bibinfo {author} {\bibfnamefont {K.~M.}\ \bibnamefont {Ranjith}},
		\bibinfo {author} {\bibfnamefont {J.}~\bibnamefont {Sichelschmidt}}, \bibinfo
		{author} {\bibfnamefont {M.}~\bibnamefont {Baenitz}}, \bibinfo {author}
		{\bibfnamefont {Y.}~\bibnamefont {Skourski}}, \bibinfo {author}
		{\bibfnamefont {F.}~\bibnamefont {Alet}}, \bibinfo {author} {\bibfnamefont
			{I.}~\bibnamefont {Rousochatzakis}}, \ and\ \bibinfo {author} {\bibfnamefont
			{A.~A.}\ \bibnamefont {Tsirlin}},\ }\bibfield  {title} {\enquote {\bibinfo
			{title} {{Hindered magnetic order from mixed dimensionalities in
					${\text{CuP}}_{2}{\text{O}}_{6}$}},}\ }\href {\doibase
		10.1103/PhysRevB.89.014407} {\bibfield  {journal} {\bibinfo  {journal} {Phys.
				Rev. B}\ }\textbf {\bibinfo {volume} {89}},\ \bibinfo {pages} {014407}
		(\bibinfo {year} {2014})}\BibitemShut {NoStop}%
	\bibitem [{\citenamefont {Ivanshin}\ \emph {et~al.}(2003)\citenamefont
		{Ivanshin}, \citenamefont {Yushankhai}, \citenamefont {Sichelschmidt},
		\citenamefont {Zakharov}, \citenamefont {Kaul},\ and\ \citenamefont
		{Geibel}}]{Ivanshin064404}%
	\BibitemOpen
	\bibfield  {author} {\bibinfo {author} {\bibfnamefont {V.~A.}\ \bibnamefont
			{Ivanshin}}, \bibinfo {author} {\bibfnamefont {V.}~\bibnamefont
			{Yushankhai}}, \bibinfo {author} {\bibfnamefont {J.}~\bibnamefont
			{Sichelschmidt}}, \bibinfo {author} {\bibfnamefont {D.~V.}\ \bibnamefont
			{Zakharov}}, \bibinfo {author} {\bibfnamefont {E.~E.}\ \bibnamefont {Kaul}},
		\ and\ \bibinfo {author} {\bibfnamefont {C.}~\bibnamefont {Geibel}},\
	}\bibfield  {title} {\enquote {\bibinfo {title} {{ESR study of the
					anisotropic exchange in the quasi-one-dimensional antiferromagnet
					${\mathrm{Sr}}_{2}{\mathrm{V}}_{3}{\mathrm{O}}_{9}$}},}\ }\href {\doibase
		10.1103/PhysRevB.68.064404} {\bibfield  {journal} {\bibinfo  {journal} {Phys.
				Rev. B}\ }\textbf {\bibinfo {volume} {68}},\ \bibinfo {pages} {064404}
		(\bibinfo {year} {2003})}\BibitemShut {NoStop}%
	\bibitem [{\citenamefont {Sichelschmidt}\ \emph {et~al.}(2002)\citenamefont
		{Sichelschmidt}, \citenamefont {Baenitz}, \citenamefont {Geibel},
		\citenamefont {Steglich}, \citenamefont {Loidl},\ and\ \citenamefont
		{Otto}}]{Sichelschmidt75}%
	\BibitemOpen
	\bibfield  {author} {\bibinfo {author} {\bibfnamefont {J.}~\bibnamefont
			{Sichelschmidt}}, \bibinfo {author} {\bibfnamefont {M.}~\bibnamefont
			{Baenitz}}, \bibinfo {author} {\bibfnamefont {C.}~\bibnamefont {Geibel}},
		\bibinfo {author} {\bibfnamefont {F.}~\bibnamefont {Steglich}}, \bibinfo
		{author} {\bibfnamefont {A.}~\bibnamefont {Loidl}}, \ and\ \bibinfo {author}
		{\bibfnamefont {H.~H.}\ \bibnamefont {Otto}},\ }\bibfield  {title} {\enquote
		{\bibinfo {title} {{Quasi-one-dimensional spin chains in CuSiO$_3$: an EPR
					study}},}\ }\href {\doibase 10.1007/BF03166185} {\bibfield  {journal}
		{\bibinfo  {journal} {Appl. Magn. Reson.}\ }\textbf {\bibinfo {volume}
			{23}},\ \bibinfo {pages} {75} (\bibinfo {year} {2002})}\BibitemShut {NoStop}%
	\bibitem [{\citenamefont {Kittel}(c2005)}]{Kittelc2005}%
	\BibitemOpen
	\bibfield  {author} {\bibinfo {author} {\bibfnamefont {C.}~\bibnamefont
			{Kittel}},\ }\href@noop {} {\emph {\bibinfo {title} {{Introduction to Solid
					State Physics}}}}\ (\bibinfo  {publisher} {J. Wiley},\ \bibinfo {address}
	{Hoboken, NJ},\ \bibinfo {year} {c2005})\ \bibinfo {note}
	{500317}\BibitemShut {NoStop}%
	\bibitem [{\citenamefont {Caslin}\ \emph {et~al.}(2014)\citenamefont {Caslin},
		\citenamefont {Kremer}, \citenamefont {Razavi}, \citenamefont {Schulz},
		\citenamefont {Mu\~noz}, \citenamefont {Pertlik}, \citenamefont {Liu},
		\citenamefont {Whangbo},\ and\ \citenamefont {Law}}]{Caslin014412}%
	\BibitemOpen
	\bibfield  {author} {\bibinfo {author} {\bibfnamefont {K.}~\bibnamefont
			{Caslin}}, \bibinfo {author} {\bibfnamefont {R.~K.}\ \bibnamefont {Kremer}},
		\bibinfo {author} {\bibfnamefont {F.~S.}\ \bibnamefont {Razavi}}, \bibinfo
		{author} {\bibfnamefont {A.}~\bibnamefont {Schulz}}, \bibinfo {author}
		{\bibfnamefont {A.}~\bibnamefont {Mu\~noz}}, \bibinfo {author} {\bibfnamefont
			{F.}~\bibnamefont {Pertlik}}, \bibinfo {author} {\bibfnamefont
			{J.}~\bibnamefont {Liu}}, \bibinfo {author} {\bibfnamefont {M.-H.}\
			\bibnamefont {Whangbo}}, \ and\ \bibinfo {author} {\bibfnamefont {J.~M.}\
			\bibnamefont {Law}},\ }\bibfield  {title} {\enquote {\bibinfo {title}
			{{Characterization of the spin-$\frac{1}{2}$ linear-chain ferromagnet
					CuAs${}_{2}$O${}_{4}$}},}\ }\href {\doibase 10.1103/PhysRevB.89.014412}
	{\bibfield  {journal} {\bibinfo  {journal} {Phys. Rev. B}\ }\textbf {\bibinfo
			{volume} {89}},\ \bibinfo {pages} {014412} (\bibinfo {year}
		{2014})}\BibitemShut {NoStop}%
	\bibitem [{\citenamefont {Bernu}\ and\ \citenamefont
		{Misguich}(2001)}]{Bernu134409}%
	\BibitemOpen
	\bibfield  {author} {\bibinfo {author} {\bibfnamefont {B.}~\bibnamefont
			{Bernu}}\ and\ \bibinfo {author} {\bibfnamefont {G.}~\bibnamefont
			{Misguich}},\ }\bibfield  {title} {\enquote {\bibinfo {title} {{Specific heat
					and high-temperature series of lattice models: Interpolation scheme and
					examples on quantum spin systems in one and two dimensions}},}\ }\href
	{\doibase 10.1103/PhysRevB.63.134409} {\bibfield  {journal} {\bibinfo
			{journal} {Phys. Rev. B}\ }\textbf {\bibinfo {volume} {63}},\ \bibinfo
		{pages} {134409} (\bibinfo {year} {2001})}\BibitemShut {NoStop}%
	\bibitem [{\citenamefont {Curro}(2009)}]{Curro026502}%
	\BibitemOpen
	\bibfield  {author} {\bibinfo {author} {\bibfnamefont {N.~J.}\ \bibnamefont
			{Curro}},\ }\bibfield  {title} {\enquote {\bibinfo {title} {{Nuclear magnetic
					resonance in the heavy fermion superconductors}},}\ }\href {\doibase
		10.1088/0034-4885/72/2/026502} {\bibfield  {journal} {\bibinfo  {journal}
			{Rep. Prog. Phys.}\ }\textbf {\bibinfo {volume} {72}},\ \bibinfo {pages}
		{026502} (\bibinfo {year} {2009})}\BibitemShut {NoStop}%
	\bibitem [{\citenamefont {Slichter}(1992)}]{Slichter1992}%
	\BibitemOpen
	\bibfield  {author} {\bibinfo {author} {\bibfnamefont {C.~P.}\ \bibnamefont
			{Slichter}},\ }\href@noop {} {\emph {\bibinfo {title} {Principle of Nuclear
				Magnetic Resonance}}},\ \bibinfo {edition} {3rd}\ ed.\ (\bibinfo  {publisher}
	{Springer},\ \bibinfo {address} {New York},\ \bibinfo {year}
	{1992})\BibitemShut {NoStop}%
	\bibitem [{\citenamefont {Lang}\ \emph {et~al.}({2005})\citenamefont {Lang},
		\citenamefont {Bobroff}, \citenamefont {Alloul}, \citenamefont {Mendels},
		\citenamefont {Blanchard},\ and\ \citenamefont {Collin}}]{Lang094404}%
	\BibitemOpen
	\bibfield  {author} {\bibinfo {author} {\bibfnamefont {G.}~\bibnamefont
			{Lang}}, \bibinfo {author} {\bibfnamefont {J.}~\bibnamefont {Bobroff}},
		\bibinfo {author} {\bibfnamefont {H.}~\bibnamefont {Alloul}}, \bibinfo
		{author} {\bibfnamefont {P.}~\bibnamefont {Mendels}}, \bibinfo {author}
		{\bibfnamefont {N.}~\bibnamefont {Blanchard}}, \ and\ \bibinfo {author}
		{\bibfnamefont {G.}~\bibnamefont {Collin}},\ }\bibfield  {title} {\enquote
		{\bibinfo {title} {{Evidence of a single nonmagnetic ${\mathrm{Co}}^{3+}$
					state in the ${\mathrm{Na}}_{1}{\mathrm{CoO}}_{2}$ cobaltate}},}\ }\href
	{\doibase 10.1103/PhysRevB.72.094404} {\bibfield  {journal} {\bibinfo
			{journal} {{Phys. Rev. B}}\ }\textbf {\bibinfo {volume} {{72}}},\ \bibinfo
		{pages} {{094404}} (\bibinfo {year} {{2005}})}\BibitemShut {NoStop}%
	\bibitem [{\citenamefont {{Ranjith, K. M. and Nath, R. and Majumder, M. and
				Kasinathan, D. and Skoulatos, M. and Keller, L. and Skourski, Y. and Baenitz,
				M. and Tsirlin, A. A.}}(2016)}]{Ranjith014415}%
	\BibitemOpen
	\bibfield  {author} {\bibinfo {author} {\bibnamefont {{Ranjith, K. M. and
					Nath, R. and Majumder, M. and Kasinathan, D. and Skoulatos, M. and Keller, L.
					and Skourski, Y. and Baenitz, M. and Tsirlin, A. A.}}},\ }\bibfield  {title}
	{\enquote {\bibinfo {title} {{Commensurate and incommensurate magnetic order
					in spin-1 chains stacked on the triangular lattice in
					${\mathrm{Li}}_{2}{\mathrm{NiW}}_{2}{\mathrm{O}}_{8}$}},}\ }\href {\doibase
		10.1103/PhysRevB.94.014415} {\bibfield  {journal} {\bibinfo  {journal} {Phys.
				Rev. B}\ }\textbf {\bibinfo {volume} {94}},\ \bibinfo {pages} {014415}
		(\bibinfo {year} {2016})}\BibitemShut {NoStop}%
	\bibitem [{\citenamefont {{M. I. Gordon and M. J. R.
				Hoch}}({1978})}]{Gordon783}%
	\BibitemOpen
	\bibfield  {author} {\bibinfo {author} {\bibnamefont {{M. I. Gordon and M. J.
					R. Hoch}}},\ }\bibfield  {title} {\enquote {\bibinfo {title} {{Quadrupolar
					spin-lattice relaxation in solids}},}\ }\href {\doibase
		10.1088/0022-3719/11/4/023} {\bibfield  {journal} {\bibinfo  {journal} {{J.
					Phys. C}}\ }\textbf {\bibinfo {volume} {{11}}},\ \bibinfo {pages} {{783}}
		(\bibinfo {year} {{1978}})}\BibitemShut {NoStop}%
	\bibitem [{\citenamefont {Simmons}\ \emph {et~al.}(1962)\citenamefont
		{Simmons}, \citenamefont {O'Sullivan},\ and\ \citenamefont
		{Robinson}}]{Simmons1168}%
	\BibitemOpen
	\bibfield  {author} {\bibinfo {author} {\bibfnamefont {W.~W.}\ \bibnamefont
			{Simmons}}, \bibinfo {author} {\bibfnamefont {W.~J.}\ \bibnamefont
			{O'Sullivan}}, \ and\ \bibinfo {author} {\bibfnamefont {W.~A.}\ \bibnamefont
			{Robinson}},\ }\bibfield  {title} {\enquote {\bibinfo {title} {{Nuclear
					Spin-Lattice Relaxation in Dilute Paramagnetic Sapphire}},}\ }\href {\doibase
		10.1103/PhysRev.127.1168} {\bibfield  {journal} {\bibinfo  {journal} {Phys.
				Rev.}\ }\textbf {\bibinfo {volume} {127}},\ \bibinfo {pages} {1168} (\bibinfo
		{year} {1962})}\BibitemShut {NoStop}%
	\bibitem [{\citenamefont {Moriya}(1956)}]{Moriya23}%
	\BibitemOpen
	\bibfield  {author} {\bibinfo {author} {\bibfnamefont {T.}~\bibnamefont
			{Moriya}},\ }\bibfield  {title} {\enquote {\bibinfo {title} {{Nuclear
					Magnetic Relaxation in Antiferromagnetics}},}\ }\href {\doibase
		10.1143/PTP.16.23} {\bibfield  {journal} {\bibinfo  {journal} {Prog. Theor.
				Exp. Phys.}\ }\textbf {\bibinfo {volume} {16}},\ \bibinfo {pages} {23--44}
		(\bibinfo {year} {1956})}\BibitemShut {NoStop}%
	\bibitem [{\citenamefont {Sandvik}(1995)}]{SandvikR9831}%
	\BibitemOpen
	\bibfield  {author} {\bibinfo {author} {\bibfnamefont {A.}~\bibnamefont
			{Sandvik}},\ }\bibfield  {title} {\enquote {\bibinfo {title} {{NMR relaxation
					rates for the spin-1/2 Heisenberg chain}},}\ }\href {\doibase
		10.1103/PhysRevB.52.R9831} {\bibfield  {journal} {\bibinfo  {journal} {Phys.
				Rev. B}\ }\textbf {\bibinfo {volume} {52}},\ \bibinfo {pages} {R9831}
		(\bibinfo {year} {1995})}\BibitemShut {NoStop}%
	\bibitem [{\citenamefont {Sachdev}(1994)}]{Sachdev13006}%
	\BibitemOpen
	\bibfield  {author} {\bibinfo {author} {\bibfnamefont {S.}~\bibnamefont
			{Sachdev}},\ }\bibfield  {title} {\enquote {\bibinfo {title} {{NMR relaxation
					in half-integer antiferromagnetic spin chains}},}\ }\href {\doibase
		10.1103/PhysRevB.50.13006} {\bibfield  {journal} {\bibinfo  {journal} {Phys.
				Rev. B}\ }\textbf {\bibinfo {volume} {50}},\ \bibinfo {pages} {13006}
		(\bibinfo {year} {1994})}\BibitemShut {NoStop}%
	\bibitem [{\citenamefont {Nath}\ \emph {et~al.}(2008)\citenamefont {Nath},
		\citenamefont {Kasinathan}, \citenamefont {Rosner}, \citenamefont {Baenitz},\
		and\ \citenamefont {Geibel}}]{Nath134451}%
	\BibitemOpen
	\bibfield  {author} {\bibinfo {author} {\bibfnamefont {R.}~\bibnamefont
			{Nath}}, \bibinfo {author} {\bibfnamefont {D.}~\bibnamefont {Kasinathan}},
		\bibinfo {author} {\bibfnamefont {H.}~\bibnamefont {Rosner}}, \bibinfo
		{author} {\bibfnamefont {M.}~\bibnamefont {Baenitz}}, \ and\ \bibinfo
		{author} {\bibfnamefont {C.}~\bibnamefont {Geibel}},\ }\bibfield  {title}
	{\enquote {\bibinfo {title} {{Electronic and magnetic properties of
					${\mathrm{K}}_{2}\mathrm{Cu}{\mathrm{P}}_{2}{\mathrm{O}}_{7}$: A model
					$S=\frac{1}{2}$ Heisenberg chain system}},}\ }\href {\doibase
		10.1103/PhysRevB.77.134451} {\bibfield  {journal} {\bibinfo  {journal} {Phys.
				Rev. B}\ }\textbf {\bibinfo {volume} {77}},\ \bibinfo {pages} {134451}
		(\bibinfo {year} {2008})}\BibitemShut {NoStop}%
	\bibitem [{\citenamefont {Lebernegg}\ \emph {et~al.}(2013)\citenamefont
		{Lebernegg}, \citenamefont {Tsirlin}, \citenamefont {Janson},\ and\
		\citenamefont {Rosner}}]{Lebernegg224406}%
	\BibitemOpen
	\bibfield  {author} {\bibinfo {author} {\bibfnamefont {S.}~\bibnamefont
			{Lebernegg}}, \bibinfo {author} {\bibfnamefont {A.~A.}\ \bibnamefont
			{Tsirlin}}, \bibinfo {author} {\bibfnamefont {O.}~\bibnamefont {Janson}}, \
		and\ \bibinfo {author} {\bibfnamefont {H.}~\bibnamefont {Rosner}},\
	}\bibfield  {title} {\enquote {\bibinfo {title} {Spin gap in malachite
				{Cu$_2$(OH)$_2$CO$_3$} and its evolution under pressure},}\ }\href {\doibase
		10.1103/PhysRevB.88.224406} {\bibfield  {journal} {\bibinfo  {journal} {Phys.
				Rev. B}\ }\textbf {\bibinfo {volume} {88}},\ \bibinfo {pages} {224406}
		(\bibinfo {year} {2013})}\BibitemShut {NoStop}%
	\bibitem [{\citenamefont {Tsirlin}\ \emph {et~al.}(2011)\citenamefont
		{Tsirlin}, \citenamefont {Nath}, \citenamefont {Sichelschmidt}, \citenamefont
		{Skourski}, \citenamefont {Geibel},\ and\ \citenamefont
		{Rosner}}]{tsirlin2011}%
	\BibitemOpen
	\bibfield  {author} {\bibinfo {author} {\bibfnamefont {A.~A.}\ \bibnamefont
			{Tsirlin}}, \bibinfo {author} {\bibfnamefont {R.}~\bibnamefont {Nath}},
		\bibinfo {author} {\bibfnamefont {J.}~\bibnamefont {Sichelschmidt}}, \bibinfo
		{author} {\bibfnamefont {Y.}~\bibnamefont {Skourski}}, \bibinfo {author}
		{\bibfnamefont {C.}~\bibnamefont {Geibel}}, \ and\ \bibinfo {author}
		{\bibfnamefont {H.}~\bibnamefont {Rosner}},\ }\bibfield  {title} {\enquote
		{\bibinfo {title} {Frustrated couplings between alternating spin-$\frac12$
				chains in {AgVOAsO$_4$}},}\ }\href {\doibase 10.1103/PhysRevB.83.144412}
	{\bibfield  {journal} {\bibinfo  {journal} {Phys. Rev. B}\ }\textbf {\bibinfo
			{volume} {83}},\ \bibinfo {pages} {144412} (\bibinfo {year}
		{2011})}\BibitemShut {NoStop}%
	\bibitem [{\citenamefont {Mukharjee}\ \emph {et~al.}(2019)\citenamefont
		{Mukharjee}, \citenamefont {Ranjith}, \citenamefont {Koo}, \citenamefont
		{Sichelschmidt}, \citenamefont {Baenitz}, \citenamefont {Skourski},
		\citenamefont {Inagaki}, \citenamefont {Furukawa}, \citenamefont {Tsirlin},\
		and\ \citenamefont {Nath}}]{mukharjee2019}%
	\BibitemOpen
	\bibfield  {author} {\bibinfo {author} {\bibfnamefont {P.~K.}\ \bibnamefont
			{Mukharjee}}, \bibinfo {author} {\bibfnamefont {K.~M.}\ \bibnamefont
			{Ranjith}}, \bibinfo {author} {\bibfnamefont {B.}~\bibnamefont {Koo}},
		\bibinfo {author} {\bibfnamefont {J.}~\bibnamefont {Sichelschmidt}}, \bibinfo
		{author} {\bibfnamefont {M.}~\bibnamefont {Baenitz}}, \bibinfo {author}
		{\bibfnamefont {Y.}~\bibnamefont {Skourski}}, \bibinfo {author}
		{\bibfnamefont {Y.}~\bibnamefont {Inagaki}}, \bibinfo {author} {\bibfnamefont
			{Y.}~\bibnamefont {Furukawa}}, \bibinfo {author} {\bibfnamefont {A.~A.}\
			\bibnamefont {Tsirlin}}, \ and\ \bibinfo {author} {\bibfnamefont
			{R.}~\bibnamefont {Nath}},\ }\bibfield  {title} {\enquote {\bibinfo {title}
			{{Bose-Einstein} condensation of triplons close to the quantum critical point
				in the quasi-one-dimensional spin-$\frac12$ antiferromagnet {NaVOPO$_4$}},}\
	}\href {\doibase 10.1103/PhysRevB.100.144433} {\bibfield  {journal} {\bibinfo
			{journal} {Phys. Rev. B}\ }\textbf {\bibinfo {volume} {100}},\ \bibinfo
		{pages} {144433} (\bibinfo {year} {2019})}\BibitemShut {NoStop}%
	\bibitem [{\citenamefont {Weickert}\ \emph {et~al.}(2019)\citenamefont
		{Weickert}, \citenamefont {Aczel}, \citenamefont {Stone}, \citenamefont
		{Garlea}, \citenamefont {Dong}, \citenamefont {Kohama}, \citenamefont
		{Movshovich}, \citenamefont {Demuer}, \citenamefont {Harrison}, \citenamefont
		{Gam{\, z}a}, \citenamefont {Steppke}, \citenamefont {Brando}, \citenamefont
		{Rosner},\ and\ \citenamefont {Tsirlin}}]{weickert2019}%
	\BibitemOpen
	\bibfield  {author} {\bibinfo {author} {\bibfnamefont {F.}~\bibnamefont
			{Weickert}}, \bibinfo {author} {\bibfnamefont {A.~A.}\ \bibnamefont {Aczel}},
		\bibinfo {author} {\bibfnamefont {M.~B.}\ \bibnamefont {Stone}}, \bibinfo
		{author} {\bibfnamefont {V.~O.}\ \bibnamefont {Garlea}}, \bibinfo {author}
		{\bibfnamefont {C.}~\bibnamefont {Dong}}, \bibinfo {author} {\bibfnamefont
			{Y.}~\bibnamefont {Kohama}}, \bibinfo {author} {\bibfnamefont
			{R.}~\bibnamefont {Movshovich}}, \bibinfo {author} {\bibfnamefont
			{A.}~\bibnamefont {Demuer}}, \bibinfo {author} {\bibfnamefont
			{N.}~\bibnamefont {Harrison}}, \bibinfo {author} {\bibfnamefont {M.~B.}\
			\bibnamefont {Gam{\, z}a}}, \bibinfo {author} {\bibfnamefont
			{A.}~\bibnamefont {Steppke}}, \bibinfo {author} {\bibfnamefont
			{M.}~\bibnamefont {Brando}}, \bibinfo {author} {\bibfnamefont
			{H.}~\bibnamefont {Rosner}}, \ and\ \bibinfo {author} {\bibfnamefont {A.~A.}\
			\bibnamefont {Tsirlin}},\ }\bibfield  {title} {\enquote {\bibinfo {title}
			{Field-induced double dome and {Bose-Einstein} condensation in the crossing
				quantum spin chain system {AgVOAsO$_4$}},}\ }\href {\doibase
		10.1103/PhysRevB.100.104422} {\bibfield  {journal} {\bibinfo  {journal}
			{Phys. Rev. B}\ }\textbf {\bibinfo {volume} {100}},\ \bibinfo {pages}
		{104422} (\bibinfo {year} {2019})}\BibitemShut {NoStop}%
\end{thebibliography}
%
	
\end{document}